\documentclass[12pt]{article}

\usepackage[margin=1in]{geometry}
\usepackage{setspace}
\usepackage{amsmath, amssymb}
\usepackage{mathtools}

\usepackage[T1]{fontenc}
\usepackage[utf8]{inputenc}

\usepackage{natbib}
\setcitestyle{numbers,square,comma}   

\usepackage{hyperref}
\hypersetup{
  colorlinks = true,
  urlcolor   = blue,
  citecolor  = blue,
  linkcolor  = black,
  pdfborder  = {0 0 0}
}
\usepackage{doi}   

\usepackage{booktabs}
\usepackage{array}
\usepackage{longtable}
\usepackage{multirow}
\usepackage{pdflscape}
\usepackage{caption}
\usepackage{graphicx}
\usepackage{float}

\usepackage{titlesec}
\titleformat{\section}{\large\bfseries}{\thesection.}{0.5em}{}
\titleformat{\subsection}{\normalsize\bfseries}{\thesubsection}{0.5em}{}
\titleformat{\subsubsection}{\normalsize\bfseries\itshape}{\thesubsubsection}{0.5em}{}

\usepackage{ragged2e}
\usepackage{enumitem}
\usepackage{afterpage}
\renewenvironment{abstract}{%
  \noindent{\bfseries Abstract}\par\noindent
  \begin{justify}\ignorespaces
}{%
  \end{justify}
}

\newcommand{\pkg}[1]{\texttt{#1}}

\begin{document}

\begin{titlepage}
\begin{center}
\vspace*{2cm}

{\Large\bfseries Multiple-group (Controlled) Interrupted Time Series Analysis with Higher-Order Autoregressive Errors: A Simulation Study Comparing Newey--West and Prais--Winsten Methods\par}

\vspace{1cm}

{\normalsize\bfseries Ariel Linden, DrPH\par}

{\small
University of California, San Francisco\\
Department of Medicine\\
Division of Clinical Informatics \& Digital Transformation (DoC-IT)\\[6pt]
\href{mailto:ariel.linden@ucsf.edu}{ariel.linden@ucsf.edu}
}

\end{center}
\end{titlepage}

\begin{abstract}
\noindent
\textbf{Background:} Previous comparisons of ordinary least squares with Newey–West standard errors (OLS-NW) and Prais–Winsten (PW) regression in multiple-group (controlled) interrupted time series analysis have been limited to first-order autoregressive (AR[1]) errors because PW estimation for higher-order AR[k] processes was not previously available. The recent extension of the Prais–Winsten transformation to AR[k] errors now permits systematic evaluation of these estimators under higher-order autoregressive dependence.

\medskip\noindent
\textbf{Methods:} A Monte Carlo simulation study based on the standard MG-ITSA regression model conducted the first systematic evaluation of OLS-NW and PW under previously unexamined AR[2] and AR[3] error structures. Simulations examined mild positive, oscillatory, and high persistent autocorrelation across varying series lengths and effect sizes. Treatment effects were defined as difference-in-differences in level and trend. Six performance measures were evaluated: power, 95\% confidence interval coverage, Type I error rate, percentage bias, root mean squared error, and empirical standard errors. Sensitivity and misspecification analyses, along with an applied example, examined the robustness and practical implications of estimator choice.

\medskip\noindent
\textbf{Results:} Both methods produced approximately unbiased estimates across all conditions. OLS-NW generally had higher power but inflated Type I error and poor coverage, especially with increasing AR order and series length. Under high persistent autocorrelation, OLS-NW coverage decreased with series length, dropping to 45–50\% at 100 periods, while PW maintained near-nominal coverage (91–94\%). OLS-NW Type I error also increased with series length, reaching 50–57\% at 100 periods, while PW remained near-nominal throughout. Under specific conditions, PW showed power advantages over OLS-NW. Sensitivity analyses supported the primary findings. Misspecification of the AR order had negligible impact on inferential performance for either method.

\medskip\noindent
\textbf{Conclusions:} The inferential trade-off previously documented under AR[1] errors not only persists but intensifies under higher-order autoregressive dependence, with OLS-NW performance deteriorating as AR order and series length increase. The power advantage of OLS-NW is mainly due to inflated false positives, especially under high persistent autocorrelation, where its inferential performance worsens with increasing series length. PW offers more reliable inference and is the preferred estimator for valid hypothesis testing, coverage, and Type I error control.

\medskip\noindent
\textbf{Keywords:} Interrupted time series analysis; multiple-group interrupted
time series analysis; controlled interrupted time series analysis;
autocorrelation; higher-order autoregression; Newey-West; Prais-Winsten
regression; power; simulation study
\end{abstract}

\newpage

\section{Introduction}

Interrupted time series analysis (ITSA) is widely used in healthcare research to evaluate the effects of interventions, policy changes, and natural events \citep{campbell1966,shadish2002}. Outcomes are observed sequentially at regular intervals, often as aggregated measures such as rates or expenditures. A single unit (e.g., a hospital or geographic region) is exposed to a treatment expected to produce an observable interruption in the level or trend of the outcome, and the design's reliance on multiple pre- and post-treatment observations gives it strong internal validity, including the ability to mitigate regression to the mean \citep{campbell1966,shadish2002,linden2013}. When applied to population-level data or replicated across settings, ITSA may also support broader external validity \citep{campbell1966,shadish2002}. The single-group design (SG-ITSA), however, remains susceptible to contemporaneous external influences and may yield misleading conclusions when outcome trends are already changing before the treatment \citep{linden2016,linden2015,linden2017a}.

These limitations can be addressed by incorporating one or more comparison
units that resemble the treated unit with respect to baseline outcome levels, pre-treatment trends, and other observable characteristics
\citep{linden2017b,abadie2010,linden2018}. This
approach, commonly referred to as multiple-group or controlled interrupted time series analysis (MG-ITSA), strengthens causal inference by providing a
comparison series that approximates the counterfactual outcome trajectory that would have been observed in the absence of the treatment \citep{rubin1974}.

Although a wide range of modeling approaches has been proposed for evaluating treatment effects in SG-ITSA, the options available for MG-ITSA are more limited due to its between-group contrast structure \citep{linden2015,linden2017c}. The primary options are ordinary least squares with Newey--West standard errors (OLS-NW) \citep{newey1987}, Prais--Winsten (PW) regression \citep{prais1954}, and autoregressive conditional heteroskedasticity (ARCH) family models \citep{harvey1989,enders2004}. However, the methodological complexity of ARCH-family models renders OLS-NW and PW the most commonly used in practice. Both handle autoregressive serial dependence, but in different ways. OLS-NW applies a post-estimation variance correction that makes no assumption about the form of autocorrelation, while PW applies a data transformation that explicitly models the AR[k] error structure. Prior comparisons have focused on AR[1] errors \citep{turner2021,bottomley2023,linden2026a}, but the recent extension of PW to AR[k] processes by Vougas \citep{vougas2021}, implemented in Stata through the \pkg{praisk} package \citep{linden2026b}, now makes it possible to evaluate both methods under higher-order autoregressive structures.

To date, no systematic evaluation has examined whether the apparent power advantages previously observed for OLS-NW under AR[1] errors persist under higher-order autoregressive dependence, or whether such gains reflect genuine statistical efficiency or inflated false positive rates. In this study, we evaluate the performance of OLS-NW and PW under AR[2] and AR[3] errors within a Monte Carlo simulation framework based on the standard MG-ITSA model. The two methods are compared across a range of design conditions, varying series length, effect size, and treatment effect type (difference-in-differences in level or trend). Data are generated under three autocorrelation patterns (mild positive, oscillatory, and high persistent) chosen to reflect plausible conditions in applied healthcare time series. In addition to statistical power, we evaluate confidence interval coverage, Type~I error rates, percentage bias, root mean squared error, and empirical standard errors. Sensitivity analyses under alternative design specifications and a misspecification analysis examining the consequences of fitting an AR[1] model to AR[2] data are also reported. An applied example using an artificial dataset based on a prediabetes disease management study is provided to illustrate the practical consequences of estimator choice.
\section{Methods}

\subsection{The MG-ITSA model}

The standard MG-ITSA regression model assumes the following form
\citep{linden2015,linden2017c}:

\begin{align}
Y_t &= \beta_0 + \beta_1 T_t + \beta_2 X_t + \beta_3 (X_t \times T_t)
       + \beta_4 Z + \beta_5 (Z \times T_t) \notag \\
    &\quad + \beta_6 (Z \times X_t) + \beta_7 (Z \times X_t \times T_t)
       + \varepsilon_t \label{eq:mgitsa}
\end{align}

\noindent where $Y_t$ is the aggregated outcome variable measured at each
time-point $t$; $T_t$ is the time since the start of the study; $X_t$ is a
dummy variable for the treatment period (0 = pre-treatment, 1 = post-treatment); $Z$ is a dummy variable denoting cohort assignment (treatment or control); and $(X_t \times T_t)$, $(Z \times T_t)$, $(Z \times X_t)$, and
$(Z \times X_t \times T_t)$ are interaction terms. Coefficients
$\beta_0$--$\beta_3$ represent the control group and $\beta_4$--$\beta_7$
represent the treatment unit. More specifically: $\beta_0$ is the intercept
(starting level) for the control group; $\beta_1$ is the pre-treatment slope
for the control group; $\beta_2$ represents the change in level for the control group immediately following introduction of the treatment; $\beta_3$ is the post-treatment change in slope for the control group; $\beta_4$ is the difference in baseline level between the treatment unit and controls; $\beta_5$ is the difference in baseline slope; $\beta_6$ indicates the difference-in-differences in level immediately following introduction of the treatment; and $\beta_7$ represents the difference-in-differences in trend after introduction of the treatment relative to pre-treatment \citep{linden2015}.

When the random error terms follow an AR[1] process,

\begin{equation}
\varepsilon_t = \rho\,\varepsilon_{t-1} + u_t \label{eq:ar1}
\end{equation}

\noindent where $\rho$ is the autocorrelation coefficient ($|\rho| < 1$) and
$u_t \sim \text{i.i.d.}\; N(0, \sigma^2)$. Equation~(\ref{eq:ar1}) naturally generalizes to an AR[$k$] process in which each error term depends on the previous $k$ lags:

\begin{equation}
\varepsilon_t = \rho_1\varepsilon_{t-1} + \rho_2\varepsilon_{t-2}
               + \cdots + \rho_k\varepsilon_{t-k} + u_t \label{eq:ark}
\end{equation}

\noindent where $u_t \sim \text{i.i.d.}\; N(0,\sigma^2)$ and $\rho_j$ is the
partial autocorrelation coefficient at lag $j$, for $j = 1, 2, \ldots, k$
\citep[see][for a detailed discussion]{kutner2005}.

The parameters $\beta_4$ and $\beta_5$ play a particularly important role in
establishing whether the treatment unit and controls are balanced on the level and trend of the outcome in the pre-treatment period. In an observational study where equivalence cannot be ensured, observed differences raise concerns about causal inference \citep{linden2015,linden2017c}. For a visual depiction of
these parameters, the reader is referred to Linden \citep{linden2015, linden2022}.

\subsection{Estimation approaches}

\subsubsection{Ordinary least squares regression with Newey--West standard errors}

This approach estimates the time series model using ordinary least squares
(OLS), with statistical inference based on heteroskedasticity- and
autocorrelation-consistent (HAC) standard errors following Newey and West
\citep{newey1987}. Because OLS makes no assumption about the form of serial
correlation, the NW correction is applicable regardless of whether the
underlying error process is AR[1] or higher-order, making it a flexible
inferential framework when the autocorrelation structure is uncertain or
potentially misspecified \citep{wooldridge2020}. The mathematical details
are as follows.

For the standard linear time series model:

\begin{equation}
    y_t = \mathbf{x}_t^{\prime}\boldsymbol{\beta} + \epsilon_t,
    \quad t = 1, 2, \ldots, T
\end{equation}

\noindent the OLS estimator is:

\begin{equation}
    \hat{\boldsymbol{\beta}} =
    (\mathbf{X}^{\prime}\mathbf{X})^{-1}\mathbf{X}^{\prime}\mathbf{y}
\end{equation}

\noindent The asymptotic distribution of the OLS estimator is:

\begin{equation}
    \sqrt{T}\,(\hat{\boldsymbol{\beta}} - \boldsymbol{\beta})
    \xrightarrow{d}
    N\!\left(\mathbf{0},\ \mathbf{Q}^{-1}\mathbf{S}\,\mathbf{Q}^{-1}\right)
\end{equation}

\noindent where $\mathbf{Q} = \text{plim}\,\frac{1}{T}\mathbf{X}^{\prime}\mathbf{X}$
is the probability limit of the regressor second-moment matrix, and $\mathbf{S}$ is
the long-run covariance matrix of the score $\mathbf{x}_t \epsilon_t$, defined as:

\begin{equation}
    \mathbf{S} = \sum_{j=-\infty}^{\infty} \boldsymbol{\Gamma}(j),
    \qquad
    \boldsymbol{\Gamma}(j) =
    E\!\left[\mathbf{x}_t\,\epsilon_t\,\epsilon_{t-j}\,
    \mathbf{x}_{t-j}^{\prime}\right]
\end{equation}

\noindent Conventional OLS assumes $\mathbf{S} = \sigma^2\mathbf{Q}$, which holds
only under homoskedasticity and no serial correlation. When either condition is
violated, OLS standard errors are inconsistent, producing unreliable $t$-statistics
and hypothesis tests.

\subsubsection{The Newey--West Estimator}

Newey and West \citep{newey1987} estimate $\mathbf{S}$ directly using a weighted sum of sample
autocovariance matrices:

\begin{equation}
    \hat{\mathbf{S}}_{NW} = \hat{\boldsymbol{\Gamma}}(0)
    + \sum_{j=1}^{L} w_j
    \left[\hat{\boldsymbol{\Gamma}}(j) +
    \hat{\boldsymbol{\Gamma}}(j)^{\prime}\right]
\end{equation}

\noindent where the sample autocovariance matrices are computed from the OLS
residuals $\hat{\epsilon}_t = y_t - \mathbf{x}_t^{\prime}\hat{\boldsymbol{\beta}}$
as:

\begin{equation}
    \hat{\boldsymbol{\Gamma}}(j)
    = \frac{1}{T}\sum_{t=j+1}^{T}
      \hat{\epsilon}_t\,\hat{\epsilon}_{t-j}\,
      \mathbf{x}_t\,\mathbf{x}_{t-j}^{\prime}
\end{equation}

\subsubsection{The Bartlett Kernel Weights}

The weights $w_j$ follow the Bartlett (triangular) kernel:

\begin{equation}
    w_j = 1 - \frac{j}{L+1}, \quad j = 1, 2, \ldots, L
\end{equation}

\noindent Weights decline linearly from near unity at lag~1 to near zero at lag~$L$,
giving less influence to higher-order autocovariances that are estimated with less
precision. This tapering scheme guarantees that $\hat{\mathbf{S}}_{NW}$ is positive
semi-definite, a necessary condition for a valid covariance matrix.

\subsubsection{Bandwidth Selection}

The lag truncation parameter $L$ controls how many lags of autocovariance are
incorporated. The data-driven rule of Newey and West \citep{newey1994} is:

\begin{equation}
    L = \left\lfloor 4 \left(\frac{T}{100}\right)^{2/9} \right\rfloor
\end{equation}

\noindent This bandwidth grows slowly with sample size $T$, balancing the bias from
omitting higher-order autocovariances against the variance introduced by estimating
too many.

\subsubsection{The Newey--West Covariance Matrix}

The full HAC covariance matrix estimator for $\hat{\boldsymbol{\beta}}$ is:

\begin{equation}
    \widehat{\mathrm{Var}}_{NW}(\hat{\boldsymbol{\beta}})
    = \left(\mathbf{X}^{\prime}\mathbf{X}\right)^{-1}
      \hat{\mathbf{S}}_{NW}
      \left(\mathbf{X}^{\prime}\mathbf{X}\right)^{-1}
\end{equation}

\noindent The Newey--West standard error for each coefficient $\hat{\beta}_k$ is the
square root of the corresponding diagonal element of
$\widehat{\mathrm{Var}}_{NW}(\hat{\boldsymbol{\beta}})$.
Coefficient estimates are unchanged; only the standard errors, $t$-statistics, and
confidence intervals are affected. Provided $L$ grows at an appropriate rate with
$T$, the resulting test statistics are asymptotically valid under heteroskedasticity
and serial correlation of unknown form \citep{wooldridge2020}.

\subsubsection{Prais--Winsten regression}

The Prais--Winsten (PW) method is a feasible generalized least squares (FGLS) procedure for linear time-series regression models originally developed to correct for AR[1] serial correlation while retaining the first observation \citep{prais1954}. In contrast to OLS-NW, PW explicitly models the autoregressive error structure and applies an exact GLS transformation to remove it, yielding coefficient estimates that are asymptotically more efficient than standard OLS when the AR specification is correct \citep{wooldridge2020}. Vougas \citep{vougas2021} extends this approach to AR[$k$] error processes of arbitrary order, implemented in Stata through the \pkg{praisk} package \citep{linden2026b}. The mathematical details are as follows.

\subsubsection{The Statistical Model}

The model consists of a linear regression equation with autoregressive errors.
The following notation is used throughout. $k$ denotes the AR lag order. $\mathbf{p}$
denotes the $k \times 1$ vector of AR parameters $(p_1, \ldots, p_k)'$; when $k = 1$
the single element is written as the scalar $p$. $q$ denotes the number of regression
coefficients including the constant. $\mathbf{L}$ denotes the $n \times n$
Prais--Winsten transformation matrix, and $\mathbf{L}_0$ its $k \times k$
initialization block.

Let $y_t$ denote the dependent variable at time $t$ and $\mathbf{x}_t$ a
$(q \times 1)$ vector of regressors. The model is:
\begin{align}
  y_t &= \mathbf{x}_t^{\prime} \boldsymbol{\beta} + u_t \label{eq:pw_reg} \\
  u_t &= p_1 u_{t-1} + p_2 u_{t-2} + \cdots + p_k u_{t-k} + \varepsilon_t
         \label{eq:pw_ar}
\end{align}
where $\varepsilon_t \sim \mathrm{iid}(0,\sigma^2_\varepsilon)$. The regression
coefficients $\boldsymbol{\beta}$ and the AR parameter vector $\mathbf{p}$ are
estimated jointly by the iterated GLS algorithm described below
\citep{judge1985,hamilton1994}.

\subsubsection{Model Specification by AR Order}

The AR(1) case follows Prais and Winsten \citep{prais1954}; AR(2) and AR($k > 2$) extend the
framework following Vougas \citep{vougas2021}.

\paragraph*{AR(1).}
With $k = 1$, stationarity requires $|p| < 1$, under which
$\sigma^2_u = \sigma^2_\varepsilon / (1-p^2)$ and the lag-$\ell$ autocovariance is
$\gamma(\ell) = p^\ell \sigma^2_u$. For $t = 2, \ldots, n$ the quasi-differenced
equation is:
\begin{equation}
  y_t - p\, y_{t-1} =
  (\mathbf{x}_t - p\, \mathbf{x}_{t-1})^{\prime} \boldsymbol{\beta} + \varepsilon_t
\end{equation}
The first observation is retained by premultiplying by $\sqrt{1-p^2}$
\citep{prais1954}:
\begin{equation}
  \tilde{y}_1 = \sqrt{1-p^2}\, y_1,
  \qquad
  \tilde{\mathbf{x}}_1 = \sqrt{1-p^2}\, \mathbf{x}_1
\end{equation}
This is the exact GLS transformation: $\sqrt{1-p^2}$ is the $(1,1)$ element of the
Cholesky factor of $\Omega^{-1}$, where $\Omega_{st} = p^{|s-t|}\sigma^2_u$.
The AR parameter $p$ is estimated at each iteration from the current residuals
$\hat{u}_t$ by:
\begin{equation}
  \hat{p} = \frac{\sum_{t=2}^{n} \hat{u}_t \hat{u}_{t-1}}
                 {\sum_{t=2}^{n} \hat{u}_{t-1}^2}
\end{equation}
This is the moment estimator from the lag-1 Yule--Walker equation
$\gamma(1) = p\gamma(0)$, corresponding to equation~(5) of Vougas \citep{vougas2021}.

\paragraph*{AR(2).}
With $k = 2$, stationarity requires all eigenvalues of the companion matrix to have
modulus less than 1, equivalent to the triangle conditions $p_1 + p_2 < 1$,
$p_2 - p_1 < 1$, and $|p_2| < 1$ \citep{hamilton1994}. For $t = 3, \ldots, n$ the
AR(2) filter is $\tilde{u}_t = u_t - p_1 u_{t-1} - p_2 u_{t-2}$. The first two
observations are transformed using $\mathbf{L}_0$, the row-reversed Cholesky factor
of $V_2^{-1}$ (see below), applied to $(y_1, y_2)^{\prime}$ and
$(x_1, x_2)^{\prime}$. The $2 \times 2$ Yule--Walker normal equations are
\citep{vougas2021}:
\begin{equation}
  \begin{bmatrix}
    \sum \hat{u}^2_{t-1}            & \sum \hat{u}_{t-1}\hat{u}_{t-2} \\
    \sum \hat{u}_{t-1}\hat{u}_{t-2} & \sum \hat{u}^2_{t-2}
  \end{bmatrix}
  \begin{bmatrix} p_1 \\ p_2 \end{bmatrix}
  =
  \begin{bmatrix}
    \sum \hat{u}_t \hat{u}_{t-1} \\
    \sum \hat{u}_t \hat{u}_{t-2}
  \end{bmatrix}
\end{equation}
solved via LU decomposition \citep{golub1996}.

\paragraph*{AR($k > 2$).}
For AR orders beyond 2, the general $k \times k$ Yule--Walker system
$A\mathbf{p} = \mathbf{b}$ is:
\begin{align}
  b[j]   &= \sum_{t=j+1}^{n} \hat{u}_t \hat{u}_{t-j},
             \quad j = 1, \ldots, k \\
  A[i,j] &= \sum_{t=\max(i,j)+1}^{n} \hat{u}_{t-i} \hat{u}_{t-j},
             \quad i,j = 1, \ldots, k
\end{align}
Cross-products are pooled across all segments and solved via LU decomposition.

\subsubsection{The Iterative GLS Algorithm}

The algorithm alternates between two steps until convergence
\citep{prais1954,judge1985,vougas2021}. Starting from OLS residuals
$\hat{\mathbf{u}} = \mathbf{y} - \mathbf{X}\hat{\boldsymbol{\beta}}_{\mathrm{OLS}}$:

\textbf{Step~1.} Estimate the AR($k$) parameters $\mathbf{p}$ from the current
residuals by solving the pooled Yule--Walker normal equations.

\textbf{Step~2.} Apply the exact Prais--Winsten GLS transformation using the current
$\mathbf{p}$ and compute GLS estimates:
\begin{equation}
  \hat{\boldsymbol{\beta}} =
  (\mathbf{X}_r^{\prime}\mathbf{X}_r)^{-1} \mathbf{X}_r^{\prime} \mathbf{y}_r
\end{equation}
where $\mathbf{y}_r$ and $\mathbf{X}_r$ are the Prais--Winsten transformed dependent
variable and regressor matrix, respectively. Convergence is declared when
$\max_j |p_j^{(\mathrm{new})} - p_j^{(\mathrm{old})}| < \tau$, where $\tau$ is a
user-specified tolerance (default $10^{-6}$).

\subsubsection{The Exact Prais--Winsten GLS Transformation}

For a data segment of length $n$ with AR($k$) errors, the transformation matrix
$\mathbf{L}$ is an $n \times n$ lower-triangular matrix. The lower $n - k$ rows
apply the standard AR filter; row $t$ (for $t > k$) produces
$\tilde{u}_t = u_t - p_1 u_{t-1} - \cdots - p_k u_{t-k}$.
The top-left $k \times k$ block is set to the reversed Cholesky factor of
$V_k^{-1}$:
\begin{equation}
  \mathbf{L}_0 = \mathrm{chol}(V_k^{-1})[k{:}1,\, k{:}1]
\end{equation}
where $V_k^{-1}$ is computed analytically using the closed-form expression of
Galbraith and Galbraith \citep{galbraith1974}. For $k = 1$ this reduces to $\sqrt{1-p^2}$; for $k \geq 2$
the procedure applies to a $k \times k$ system with no special-case treatment required.

\subsubsection{AR Parameter Bounds and Stationarity}

The stationarity condition for AR($k$) is that all eigenvalues of the $k \times k$
companion matrix $C$ have modulus strictly less than 1 \citep{hamilton1994}:
\begin{equation}
  C =
  \begin{bmatrix}
    p_1    & p_2    & \cdots & p_{k-1} & p_k    \\
    1      & 0      & \cdots & 0       & 0      \\
    0      & 1      & \cdots & 0       & 0      \\
    \vdots &        & \ddots &         & \vdots \\
    0      & 0      & \cdots & 1       & 0
  \end{bmatrix}
\end{equation}
There is no constraint on the magnitude of any individual $p_j$; the stationarity
region for AR(2) is the triangle defined by $p_2 + p_1 < 1$, $p_2 - p_1 < 1$, and
$-1 < p_2 < 1$ \citep{hamilton1994}.

\subsubsection{Panel Data and Multiple Segments}

Cross-product matrices $A$ and $\mathbf{b}$ are accumulated within each contiguous
segment and summed across all segments, yielding a single pooled $\mathbf{p}$ estimate
\citep{park1980}. The Prais--Winsten transformation is applied independently to each
segment, with the exact initialization restarted at the beginning of every segment.
Cross-segment autocorrelation is assumed to be zero.

\subsubsection{GLS Coefficient Covariance and Standard Errors}

At convergence, the covariance matrix of $\hat{\boldsymbol{\beta}}$ is
\citep{judge1985}:
\begin{equation}
  \widehat{V}(\hat{\boldsymbol{\beta}}) =
  s^2 (\mathbf{X}_r^{\prime}\mathbf{X}_r)^{-1},
  \qquad
  s^2 = \frac{\mathbf{e}^{\prime}\mathbf{e}}{N - q}
\end{equation}
where $\mathbf{e} = \mathbf{y}_r - \mathbf{X}_r \hat{\boldsymbol{\beta}}$ are the
transformed residuals and $q$ is the number of active regressors including the
constant. The asymptotic covariance of the Yule--Walker estimator is
$\mathrm{Var}(\hat{\mathbf{p}}) = \hat{\sigma}^2_\varepsilon A^{-1}$
\citep{brockwell1991}, where $A$ is the pooled cross-product matrix from the final
Yule--Walker solve.

\subsection{The \pkg{power\_itsa} package for Stata}

All simulations were conducted using the community-contributed Stata package \pkg{power\_itsa} \citep{linden2025}, which computes statistical power for ITSA designs via simulation. The package uses user-specified coefficient values from the MG-ITSA regression model (Equation~\ref{eq:mgitsa}) to generate artificial
datasets through the \pkg{itsadgp} package \citep{linden2024}. This
data-generating process produces one treated unit and a user-specified number of control units. Each generated dataset is passed to the \pkg{itsa} package \citep{linden2015} to estimate treatment effects corresponding to $\beta_6$ and $\beta_7$ (the difference-in-differences in level and trend, respectively). The package supports both OLS-NW and PW regression (via \pkg{praisk}) as estimation methods, and all analyses were conducted using both approaches. Following model estimation, \pkg{power\_itsa} performs Wald tests to evaluate whether $\beta_6
= 0$ and $\beta_7 = 0$. For all analyses, the null hypothesis is rejected at $\alpha = 0.05$.

\subsection{Simulation strategy}

Table~\ref{tab:sim_inputs} presents the simulation inputs. The primary
objective was to evaluate whether, and to what extent, the magnitude and
structure of autocorrelation under AR[2] and AR[3] processes differentially
influence the performance of OLS-NW and PW estimators across a range of
conditions, including variation in series length, effect size, and effect type (difference-in-differences in level or trend).

Simulation inputs were modeled on those used in Linden \citep{linden2026a} to maintain comparability, with three key modifications. First, the number of time periods was extended to range from 10 to 100 (compared with an upper limit of 50 in \citep{linden2026a}) to assess whether higher-order autoregressive error structures exhibit different behavior over longer time horizons. Second, the number of control units was fixed at 4. This value represented both the minimum tested and the most variable condition in prior work \citep{linden2026a}, so fixing it preserves comparability while reducing simulation burden. Third,
this study introduces higher-order autoregressive error structures. For both AR[2] and AR[3], three autocorrelation scenarios were specified: mild positive autocorrelation, oscillatory autocorrelation with damped behavior, and high persistent positive autocorrelation (see Table~\ref{tab:sim_inputs} for parameter values).

The three autocorrelation scenarios were chosen to represent qualitatively distinct patterns of serial dependence that are plausible in applied healthcare time series. Scenario 1 (mild positive autocorrelation: AR[2] $\rho = (0.4, 0.2)$; AR[3] $\rho = (0.4, 0.2, 0.1)$) has a maximum companion matrix eigenvalue of $\sim$0.69 and 0.72, respectively. The autocovariance function decays monotonically, and both AR coefficients are positive, producing the kind of smooth, persistent positive correlation commonly encountered in routine healthcare data. This scenario serves as the baseline case and is most directly comparable to the AR[1] conditions examined in prior work \citep{linden2026a}. Scenario 2 (oscillatory autocorrelation: AR[2] $\rho = (0.5, -0.4)$; AR[3] $\rho = (0.7, -0.3, 0.15)$) has a maximum eigenvalue of $\sim$0.63 and 0.84, respectively. The negative second-order coefficient produces an autocovariance function that alternates in sign with a cycle of $\sim$5--6 periods, creating a complex spectral structure near frequency zero. This scenario was included to assess whether the sign-alternating autocovariance pattern differentially affects the two methods' inferential behavior relative to smooth positive autocorrelation. Scenario 3 (high persistent positive autocorrelation: AR[2] $\rho = (0.7, 0.2)$; AR[3] $\rho = (0.6, 0.25, 0.1)$) has a maximum eigenvalue of $\sim$0.90 and 0.93, respectively, both approaching but remaining within the unit circle, ensuring stationarity. The autocovariance function decays very slowly, producing near-unit-root behavior. This scenario represents the most demanding condition for both methods and most closely resembles the persistent autocorrelation patterns that can arise in cumulative or slowly evolving health system outcomes.

All remaining inputs follow Linden \citep{linden2026a,linden2026c}. The
starting level (intercept) was set to 10 for both the treated unit and
controls. The pre-treatment trend was set to 1 for both series. The change in level for the control series at the initiation of the treatment was set to 0. For the treated unit, the level change was set to 2, 2.5, and 3 when evaluating level effects (representing increases of 20\%, 25\%, and 30\% relative to the counterfactual), and to 0 when evaluating trend effects. The post-treatment trend for the control series was set to 1. For the treated unit, the post-treatment trend was set to 1.25, 1.5, and 2 when evaluating trend effects (representing increases of 25\%, 50\%, and 100\% over the pre-treatment slope), and to 0 when evaluating level effects. A standard deviation of 1 was used for the normally distributed random error term. The treatment was initiated at the midpoint of the time series in all scenarios.

\subsection{Misspecification analysis}

A supplementary simulation examined the consequences of AR order underspecification, in which data are generated under an AR[2] process but the model is estimated assuming an AR[1] structure. This scenario reflects common applied practice, where researchers may not conduct formal autocorrelation diagnostics or may default to a single-lag correction. Overspecification was not examined as it represents the less practically relevant scenario.

The data-generating process followed the same design as the primary simulations: one treated unit and four controls, a treatment initiation at the midpoint of the series, and series lengths ranging from 10 to 100 periods. Three AR[2] autocorrelation scenarios were examined: mild positive ($\rho = (0.4, 0.2)$), oscillatory ($\rho = (0.5, -0.4)$), and high persistent ($\rho = (0.7, 0.2)$). For each scenario, two estimation conditions were compared: correct AR[2] specification (DGP = AR[2], fit = AR[2]) and underspecification (DGP = AR[2], fit = AR[1]). Both OLS-NW and PW were evaluated under each condition. The post-treatment effect parameter was set to 1.5 for power simulations (representing a 50\% trend effect) and to 1 for Type~I error simulations. All other inputs followed the primary simulation design. Each condition was replicated 2,000 times. Performance was assessed using power, 95\% confidence interval coverage, Type~I error rate, percentage bias, RMSE, and empirical standard errors. 

\subsection{Sensitivity analysis}

The sensitivity analysis examined whether the combination of level and trend changes differentially affects the performance of OLS-NW and PW models under the AR[2] and AR[3] scenarios. The controls' baseline level was set to 8, and the treated unit's baseline level was set to 10. The treatment unit's level change was set to 2 (a 20\% increase), with no level change for the controls. The treatment unit's post-treatment trend was set to 2 (a 100\% increase), and the controls' post-treatment trend was set to 1, reflecting continuation of the baseline trend.

\subsection{Performance measures}

The performance of OLS-NW and PW was evaluated using six metrics: statistical power, 95\% confidence interval coverage, Type~I error rate, percentage bias, root mean squared error (RMSE), and empirical standard errors \citep{burton2006}. Across all primary, sensitivity, and misspecification analyses, 720 unique design conditions were evaluated, each with 2,000 replications, yielding a total of 1,440,000 simulated datasets. All analyses were conducted using Stata\textsuperscript{TM} version 19.0.

\section{Results}

\subsection{Difference-in-Differences in Trend: AR{[2]}}

\subsubsection{Power}

\begin{figure}[H]
  \centering
  \includegraphics[width=0.85\textwidth, height=0.85\textheight, keepaspectratio]{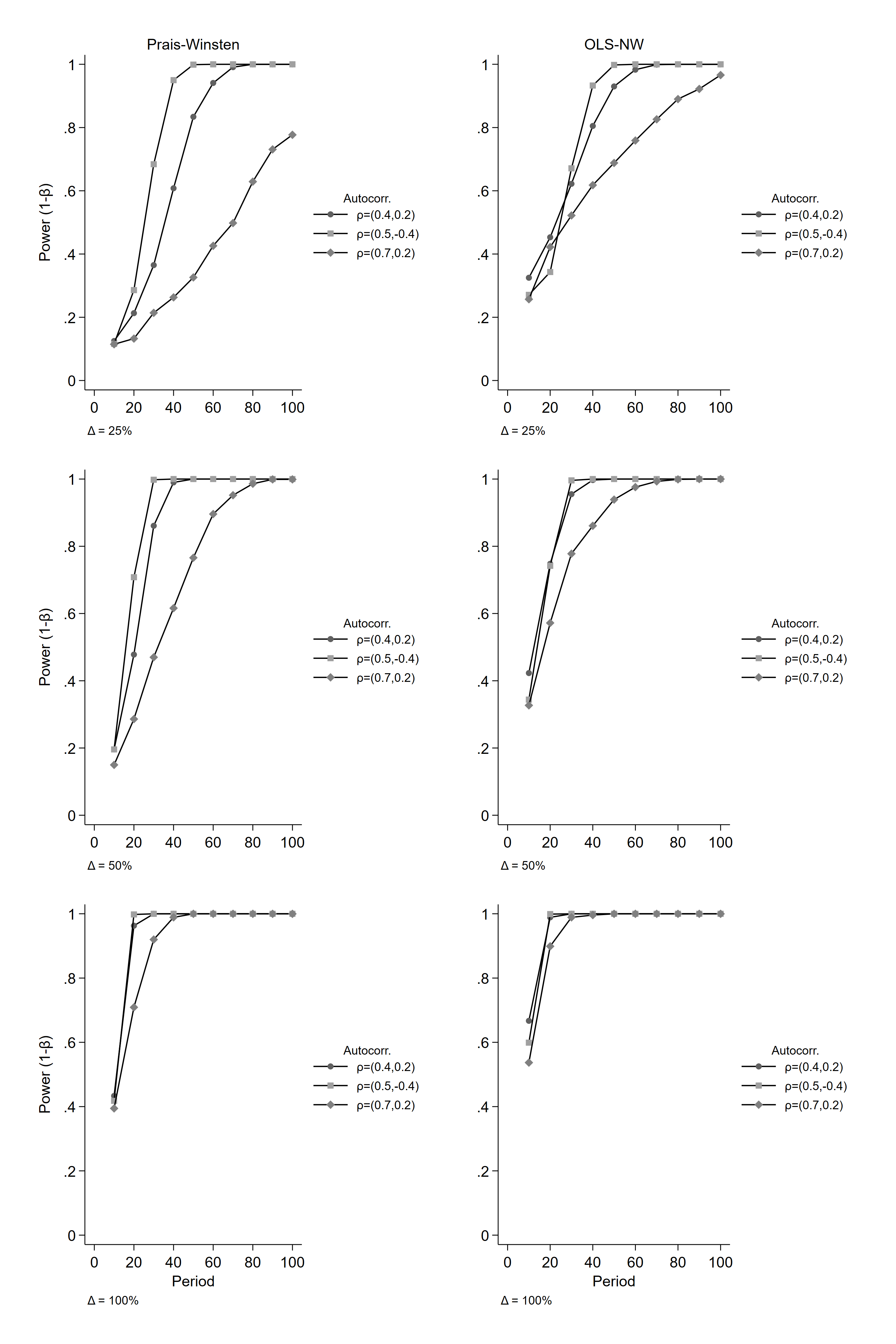}
  \caption{Power (1$-\beta$) for the difference-in-differences in trend under AR[2] error structures. Left column: Prais-Winsten; right column: OLS-NW. Rows represent effect sizes (25\%, 50\%, 100\%). Lines distinguish autocorrelation scenarios: mild positive $\rho=(0.4, 0.2)$ (circles); oscillatory $\rho=(0.5, -0.4)$ (squares); high persistent $\rho=(0.7, 0.2)$ (diamonds).}
  \label{fig:1}
\end{figure}

Figure~\ref{fig:1} presents power ($1-\beta$) for the difference-in-differences in trend under AR[2] error structures. Across all three AR[2] parameter conditions and effect sizes, OLS-NW consistently achieved higher power than PW at shorter series lengths, with both methods converging to maximum power as the number of periods increased. Under mild positive autocorrelation, OLS-NW reached near-maximum power by $\sim$40--50 periods across all effect sizes, whereas PW required somewhat longer series to reach comparable levels. Under oscillatory autocorrelation, the gap between the two methods narrowed considerably, with both converging to maximum power at similar rates. The largest differences were observed under high persistent positive autocorrelation, where PW remained substantially underpowered relative to OLS-NW across all effect sizes. At the smallest effect size (25\%), PW did not reach maximum power even at 100 periods, whereas OLS-NW achieved near-maximum power by $\sim$60--70 periods. As expected, larger effect sizes accelerated convergence to maximum power for both methods, although the relative advantage of OLS-NW persisted across all conditions. As discussed below, however, this apparent advantage reflects anticonservative inference rather than genuine gains in statistical efficiency.

\subsubsection{95\% Coverage}

\begin{figure}[H]
  \centering
  \includegraphics[width=0.85\textwidth, height=0.85\textheight, keepaspectratio]{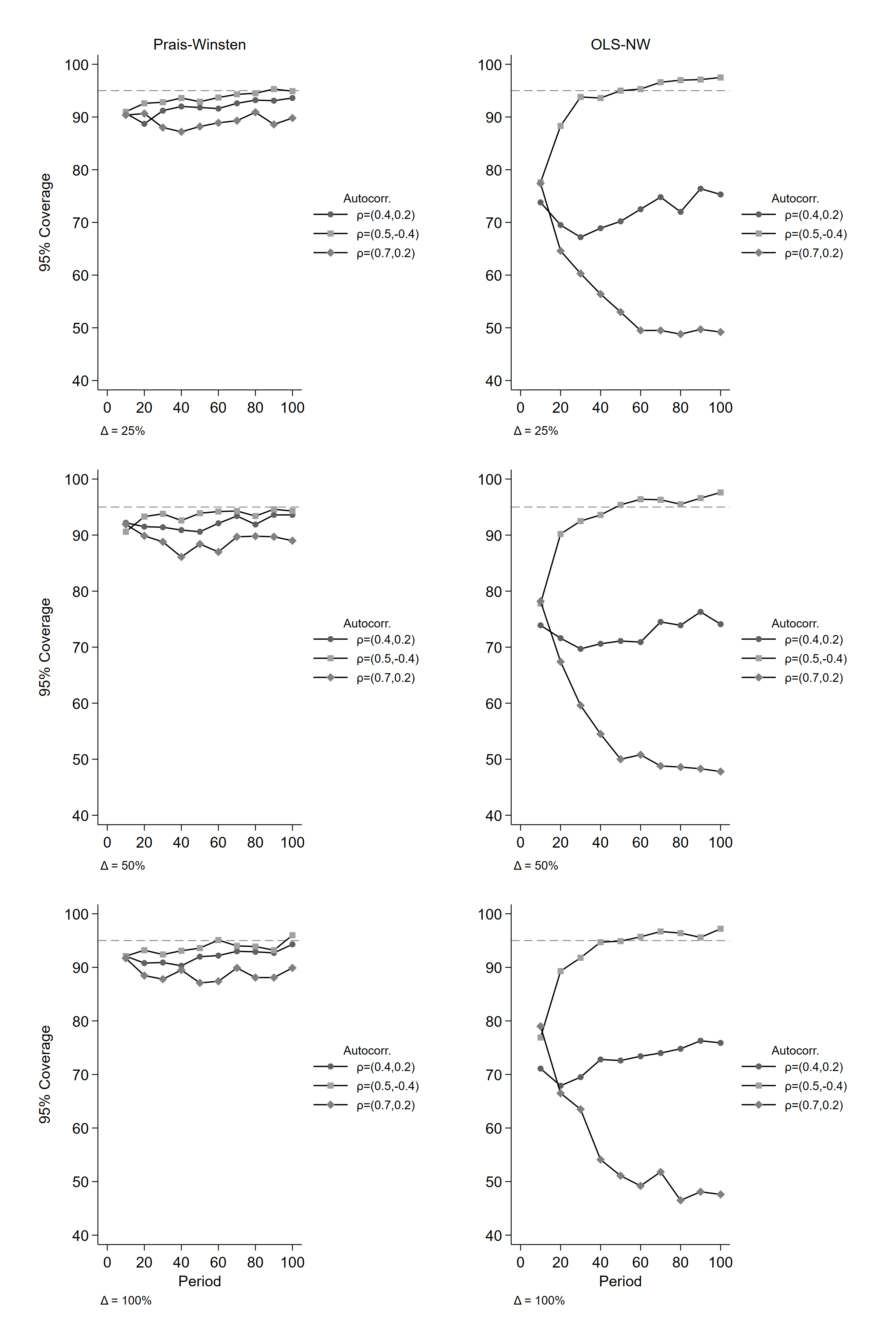}
  \caption{95\% confidence interval coverage for the difference-in-differences in trend under AR[2] error structures. Left column: Prais-Winsten; right column: OLS-NW. Rows represent effect sizes (25\%, 50\%, 100\%). Lines distinguish autocorrelation scenarios: mild positive $\rho=(0.4, 0.2)$ (circles); oscillatory $\rho=(0.5, -0.4)$ (squares); high persistent $\rho=(0.7, 0.2)$ (diamonds). Dashed reference line at nominal 95\%.}
  \label{fig:2}
\end{figure}

Figure~\ref{fig:2} presents 95\% coverage rates for the difference-in-differences in trend under AR[2] error structures. Under mild positive autocorrelation, PW maintained coverage close to the nominal 95\% level across all series lengths and effect sizes, whereas OLS-NW exhibited substantial and persistent undercoverage, remaining around 75\% with no meaningful improvement as the number of periods increased. This deficit was stable across all three effect sizes, indicating that it reflects a structural feature of OLS-NW under this autocorrelation condition rather than a small-sample artifact. Under oscillatory autocorrelation, PW maintained near-nominal coverage throughout. OLS-NW started below the nominal level at short series lengths but converged upward toward 95\% as the number of periods increased, reaching near-nominal coverage by $\sim$80--100 periods. The most consequential pattern emerged under high persistent positive autocorrelation. PW maintained stable coverage near 90\% across all series lengths and effect sizes. In contrast, OLS-NW coverage deteriorated monotonically as the number of periods increased, declining from $\sim$80\% at 10 periods to $\sim$50\%
at 100 periods. 

\subsubsection{Type I Error}

\begin{figure}[H]
  \centering
  \includegraphics[width=\textwidth, height=0.35\textheight, keepaspectratio]{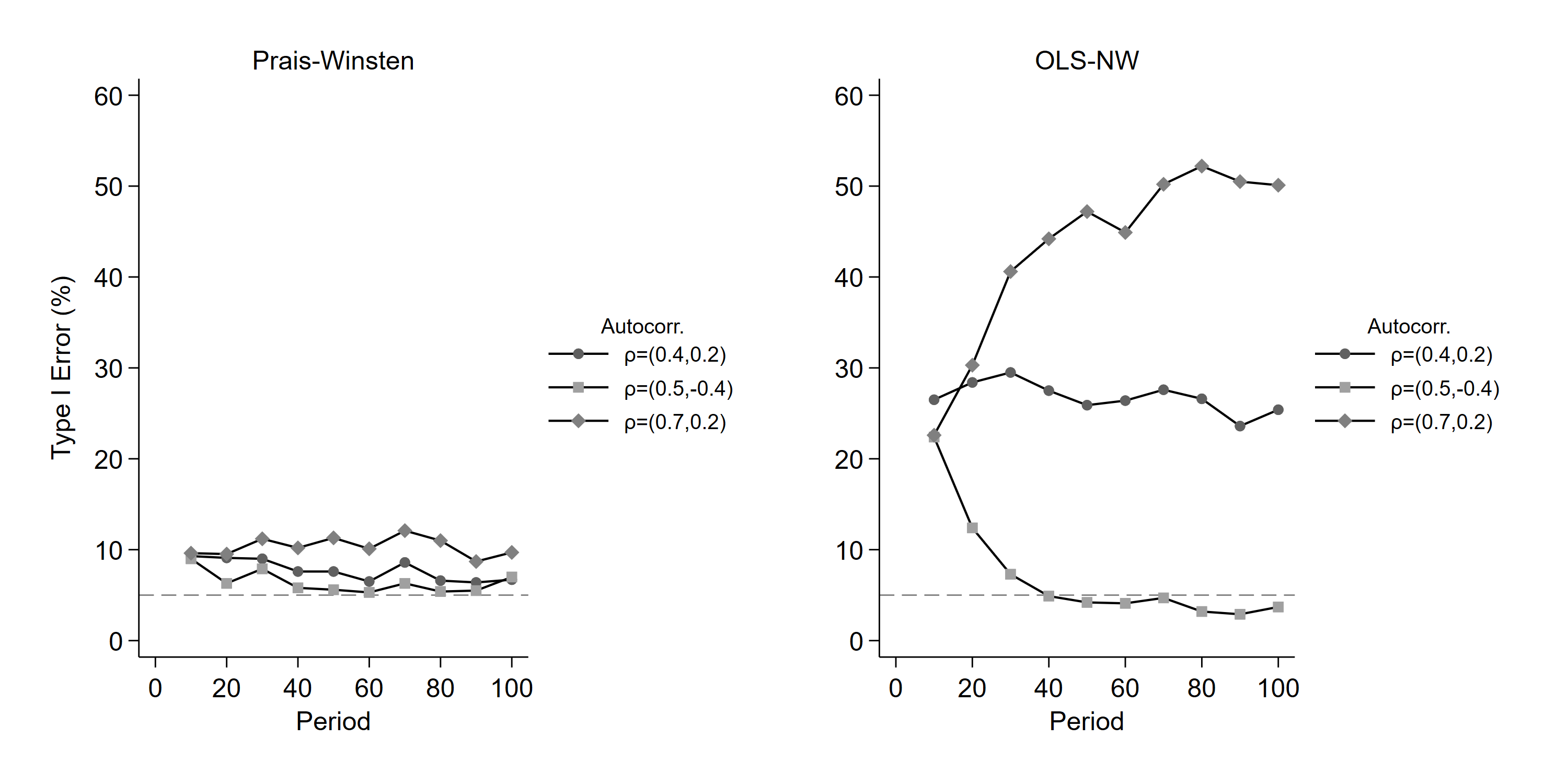}
  \caption{Type~I error rates for the difference-in-differences in trend under AR[2] error structures. Left panel: Prais-Winsten; right panel: OLS-NW. Lines distinguish autocorrelation scenarios: mild positive $\rho=(0.4, 0.2)$ (circles); oscillatory $\rho=(0.5, -0.4)$ (squares); high persistent $\rho=(0.7, 0.2)$ (diamonds). Dashed reference line at nominal 5\%.}
  \label{fig:3}
\end{figure}

Figure~\ref{fig:3} presents Type~I error rates for the difference-in-differences in trend under AR[2] error structures. PW maintained Type~I error close to the nominal 5\% level across all three AR[2] parameter conditions and the full range of series lengths, indicating well-controlled false positive rates. In contrast, OLS-NW exhibited markedly inflated Type~I error in two of the three conditions. Under mild positive autocorrelation, OLS-NW produced Type~I error rates of $\sim$25--30\% across all series lengths, with no tendency to improve as
the number of periods increased. Under oscillatory autocorrelation, Type~I error was severely inflated at short series lengths ($\sim$22\%) but declined toward the nominal level by $\sim$40 periods, suggesting that this is primarily a small-sample issue under this condition. Under high persistent positive autocorrelation, OLS-NW exhibited the most severe pattern: Type~I error increased monotonically with series length, rising from $\sim$22\% at short series lengths to $\sim$50\% at 100 periods. 

\subsubsection{Bias}

\begin{figure}[H]
  \centering
  \includegraphics[width=0.85\textwidth, height=0.85\textheight, keepaspectratio]{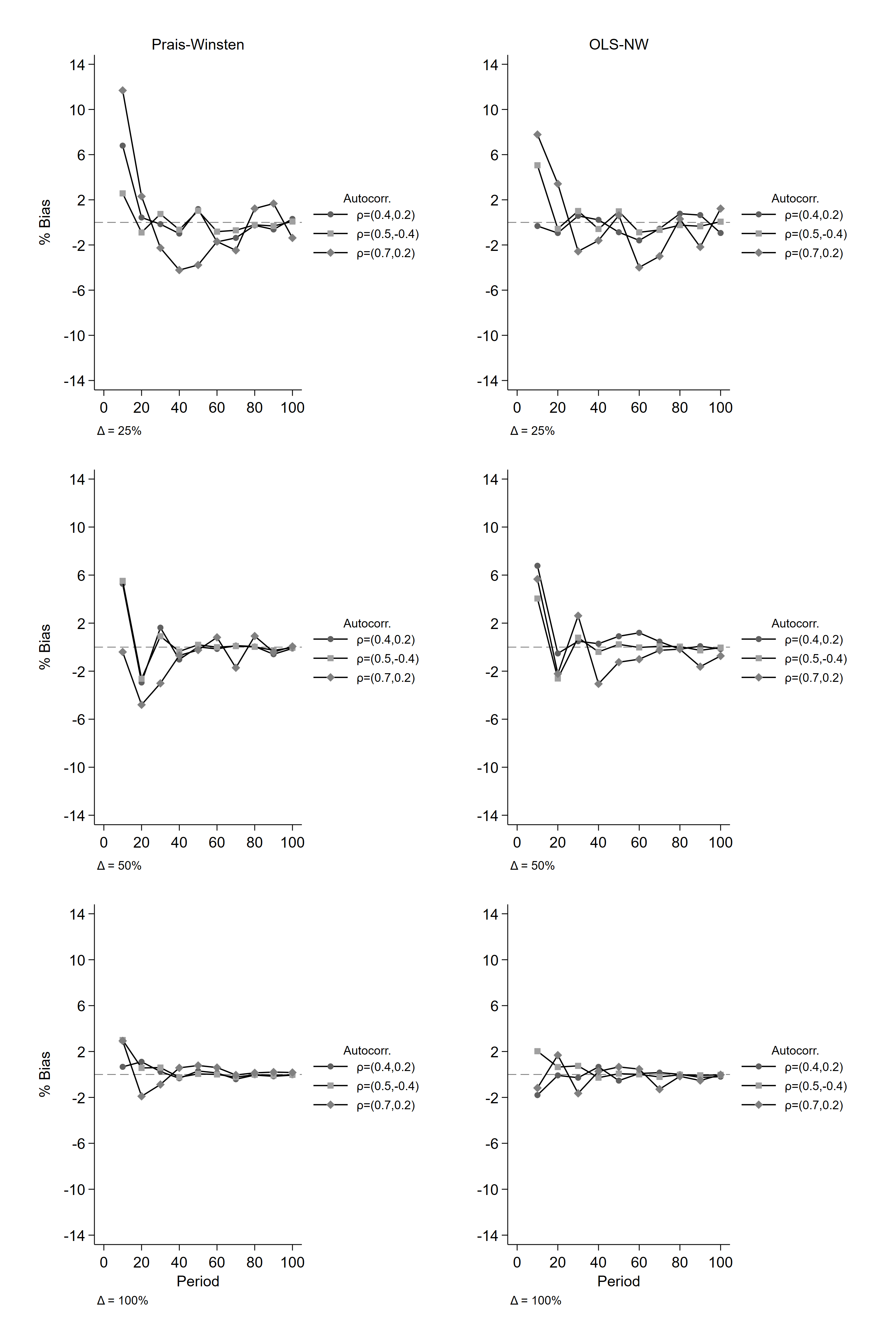}
  \caption{Percentage bias for the difference-in-differences in trend under AR[2] error structures. Left column: Prais-Winsten; right column: OLS-NW. Rows represent effect sizes (25\%, 50\%, 100\%). Lines distinguish autocorrelation scenarios: mild positive $\rho=(0.4, 0.2)$ (circles); oscillatory $\rho=(0.5, -0.4)$ (squares); high persistent $\rho=(0.7, 0.2)$ (diamonds). Dashed reference line at zero.}
  \label{fig:4}
\end{figure}

Figure~\ref{fig:4} presents percentage bias for the difference-in-differences in trend under AR[2] error structures. Both OLS-NW and PW were essentially unbiased across all autocorrelation conditions, effect sizes, and series lengths, with percentage bias remaining close to zero in nearly all scenarios and no evidence of systematic directional bias. Some instability was observed at very short series lengths (10--20 periods), particularly under oscillatory and high persistent autocorrelation, where both methods showed modest variability before stabilizing. These fluctuations diminished quickly and were not meaningfully different between methods. Overall, the bias results indicate that differences in performance between OLS-NW and PW arise from variance and inferential properties rather than systematic error in point estimation.

\subsubsection{RMSE}

\begin{figure}[H]
  \centering
  \includegraphics[width=0.85\textwidth, height=0.85\textheight, keepaspectratio]{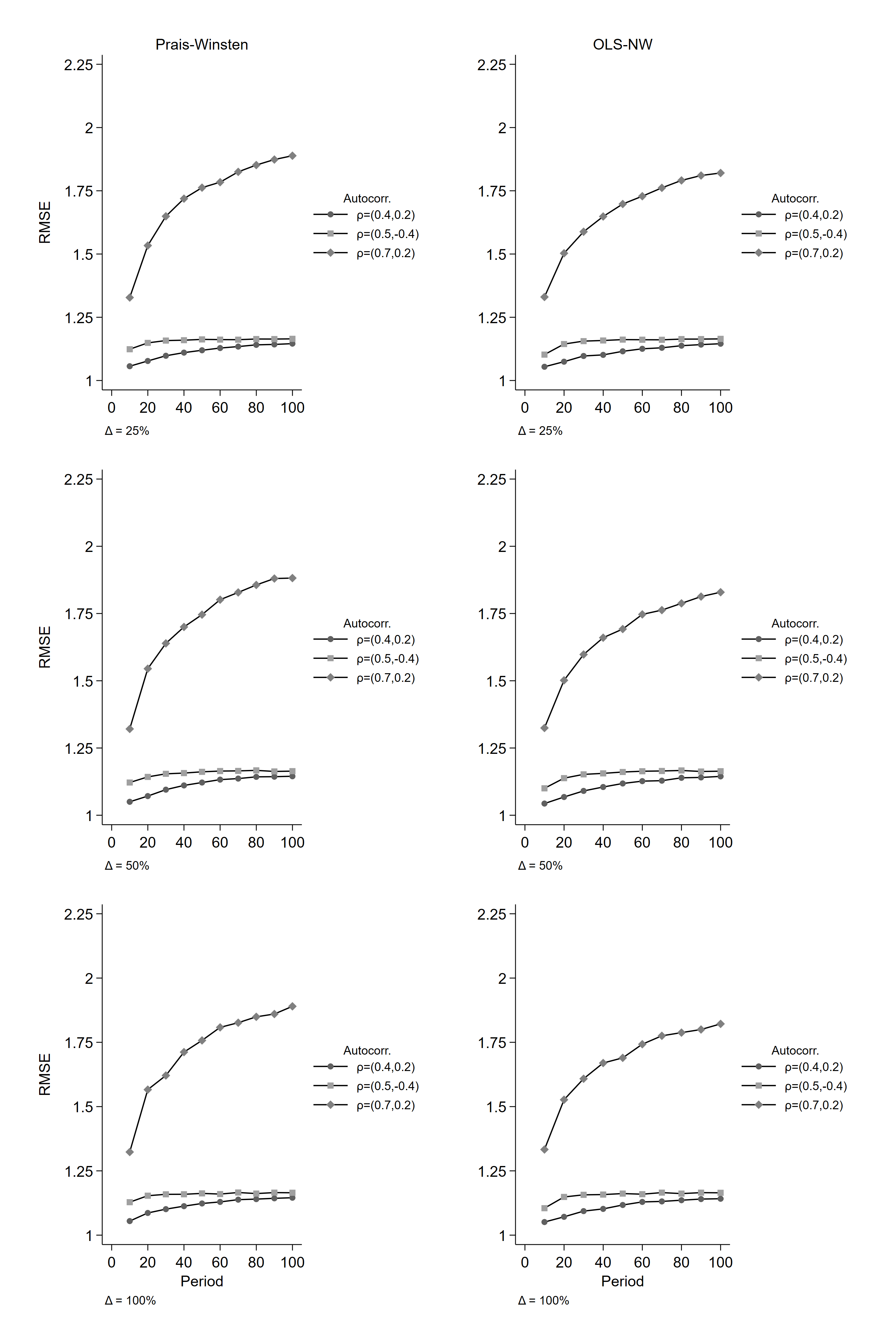}
  \caption{Root mean squared error (RMSE) for the difference-in-differences in trend under AR[2] error structures. Left column: Prais-Winsten; right column: OLS-NW. Rows represent effect sizes (25\%, 50\%, 100\%). Lines distinguish autocorrelation scenarios: mild positive $\rho=(0.4, 0.2)$ (circles); oscillatory $\rho=(0.5, -0.4)$ (squares); high persistent $\rho=(0.7, 0.2)$ (diamonds).}
  \label{fig:5}
\end{figure}

Figure~\ref{fig:5} presents root mean squared error (RMSE) for the
difference-in-differences in trend under AR[2] error structures. Under mild
positive and oscillatory autocorrelation, both methods produced closely aligned and relatively flat RMSE profiles, increasing only slightly from $\sim$1.0 at 10 periods to $\sim$1.15 at 100 periods, with negligible differences between OLS-NW and PW across all effect sizes. Under high persistent positive autocorrelation, RMSE increased substantially and monotonically for both methods, rising from $\sim$1.3 at 10 periods to $\sim$1.9 at 100 periods. In this setting, PW produced marginally higher RMSE than OLS-NW throughout, although the two methods tracked each other closely. Given the negligible bias observed across conditions, the elevated RMSE under high persistent autocorrelation reflects increased variance rather than systematic misestimation.

\subsubsection{Empirical Standard Errors}

\begin{figure}[H]
  \centering
  \includegraphics[width=0.85\textwidth, height=0.85\textheight, keepaspectratio]{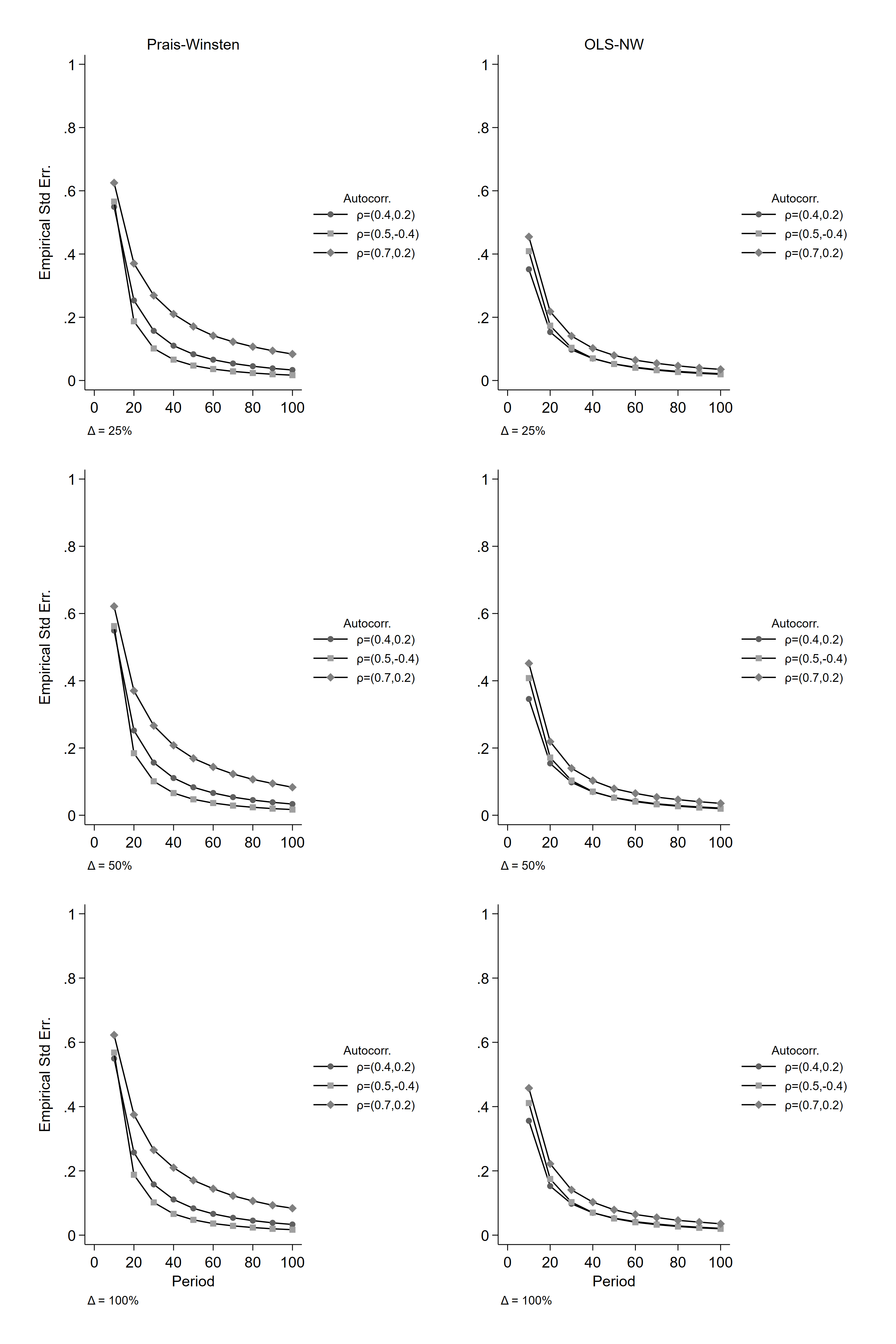}
  \caption{Empirical standard errors for the difference-in-differences in trend under AR[2] error structures. Left column: Prais-Winsten; right column: OLS-NW. Rows represent effect sizes (25\%, 50\%, 100\%). Lines distinguish autocorrelation scenarios: mild positive $\rho=(0.4, 0.2)$ (circles); oscillatory $\rho=(0.5, -0.4)$ (squares); high persistent $\rho=(0.7, 0.2)$ (diamonds).}
  \label{fig:6}
\end{figure}

Figure~\ref{fig:6} presents empirical standard errors for the
difference-in-differences in trend under AR[2] error structures. Both methods exhibited declining empirical standard errors as series length increased, consistent with improved estimation precision in larger samples. PW consistently produced larger empirical standard errors than OLS-NW across all autocorrelation conditions and effect sizes. Under mild positive and oscillatory autocorrelation, the gap between the two methods narrowed as the number of periods increased, with both converging to similar values by $\sim$40--50 periods. Under high persistent positive autocorrelation, PW maintained noticeably larger empirical standard errors than OLS-NW across the full range of series lengths, with the separation remaining evident even at 100 periods. These
differences are consistent with the broader pattern across performance measures: OLS-NW achieves smaller standard errors and higher power, but at the cost of inflated Type~I error and undercoverage, particularly under high persistent autocorrelation.

\subsection{Difference-in-Differences in Trend: AR{[3]}}

\subsubsection{Power}

\begin{figure}[H]
  \centering
  \includegraphics[width=0.85\textwidth, height=0.85\textheight, keepaspectratio]{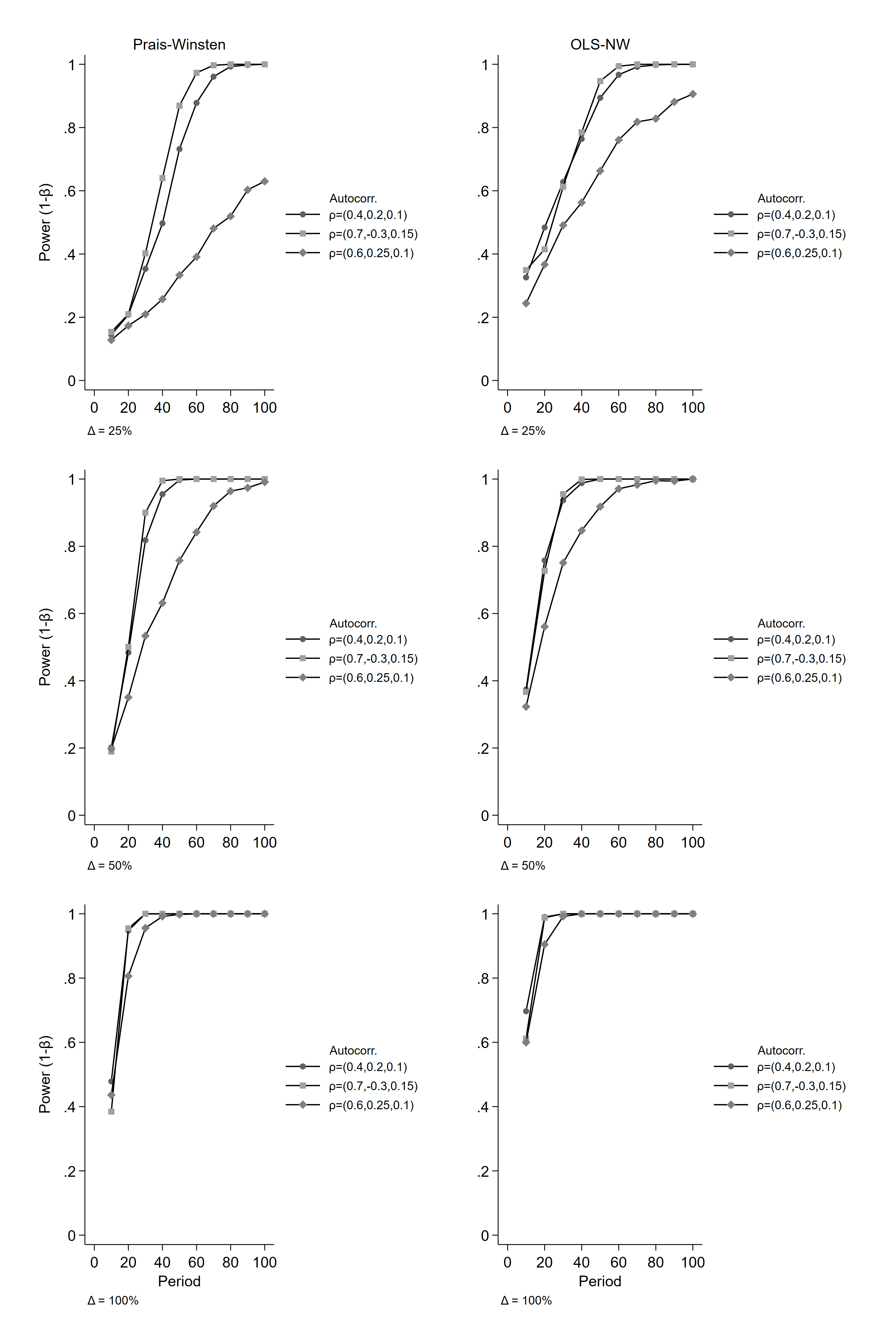}
  \caption{Power (1$-\beta$) for the difference-in-differences in trend under AR[3] error structures. Left column: Prais-Winsten; right column: OLS-NW. Rows represent effect sizes (25\%, 50\%, 100\%). Lines distinguish autocorrelation scenarios: mild positive $\rho=(0.4, 0.2, 0.1)$ (circles); oscillatory $\rho=(0.7, -0.3, 0.15)$ (squares); high persistent $\rho=(0.6, 0.25, 0.1)$ (diamonds).}
  \label{fig:7}
\end{figure}

Figure~\ref{fig:7} presents power ($1-\beta$) for the difference-in-differences in trend under AR[3] error structures. The overall pattern closely mirrors that observed under AR[2], with OLS-NW consistently achieving higher power than PW at shorter series lengths across all three AR[3] parameter conditions and effect sizes. Under mild positive autocorrelation, OLS-NW reached near-maximum power earlier than PW, although both methods converged to maximum power by $\sim$50--60 periods at the 25\% effect size and considerably sooner at larger effect sizes. Under oscillatory autocorrelation, a similar pattern was
observed, with OLS-NW maintaining a consistent lead while both methods converged to maximum power at $\sim$50--60 periods. The largest differences were observed under high persistent positive autocorrelation, where the gap between the two methods was substantially larger than under AR[2]. OLS-NW approached near-maximum power by $\sim$80--90 periods at the 25\% effect size, whereas PW reached only $\sim$60\% power at 100 periods. At larger effect sizes the gap narrowed, but PW continued to lag OLS-NW considerably, indicating that the power disadvantage of PW under high persistent autocorrelation becomes more pronounced as AR order increases from 2 to 3. As with the AR[2] results, however, OLS-NW's apparent power advantage reflects anticonservative inference---inflated Type~I error and poor coverage---rather than genuine gains in statistical efficiency.

\subsubsection{95\% Coverage}

\begin{figure}[H]
  \centering
  \includegraphics[width=0.85\textwidth, height=0.85\textheight, keepaspectratio]{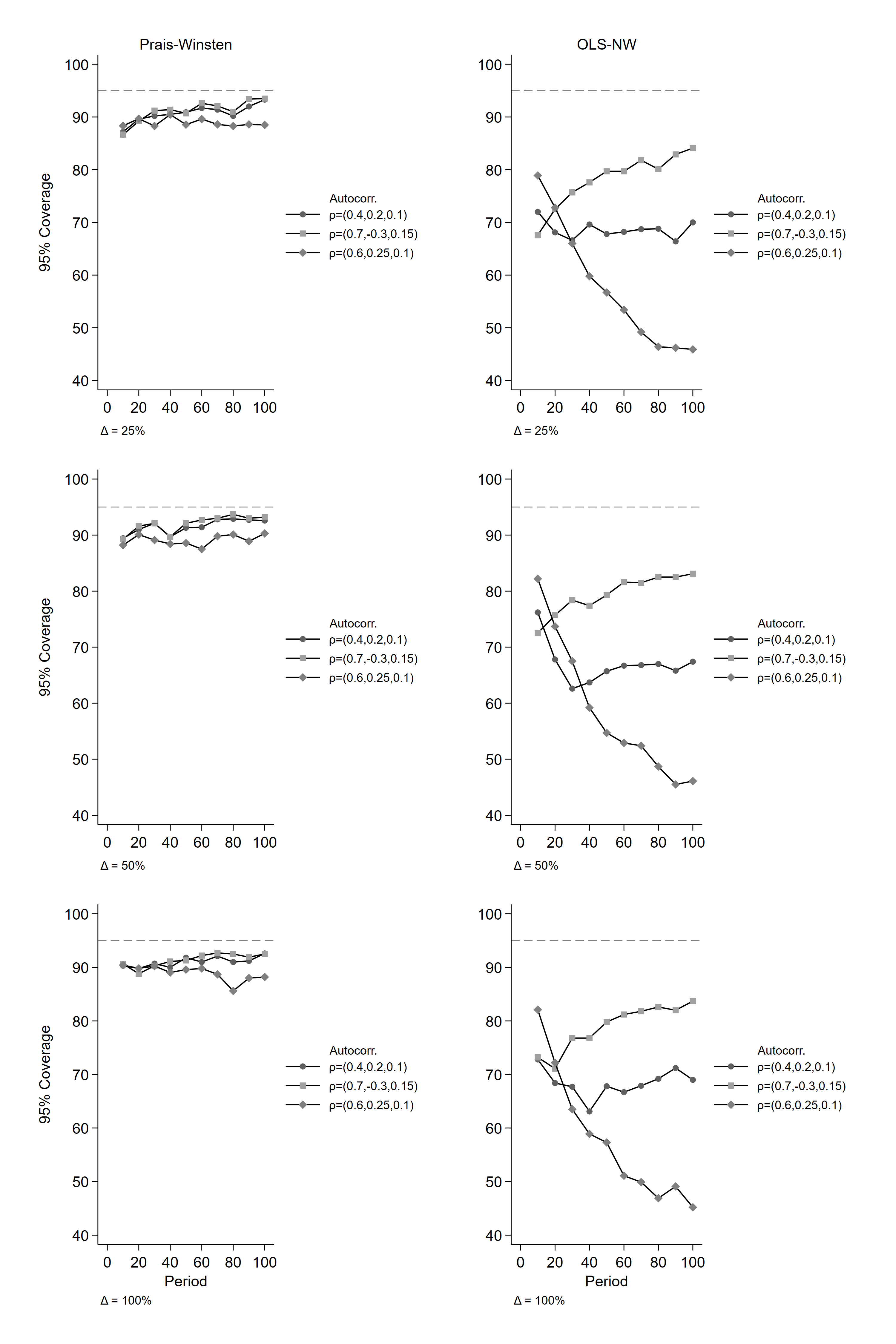}
  \caption{95\% confidence interval coverage for the difference-in-differences in trend under AR[3] error structures. Left column: Prais-Winsten; right column: OLS-NW. Rows represent effect sizes (25\%, 50\%, 100\%). Lines distinguish autocorrelation scenarios: mild positive $\rho=(0.4, 0.2, 0.1)$ (circles); oscillatory $\rho=(0.7, -0.3, 0.15)$ (squares); high persistent $\rho=(0.6, 0.25, 0.1)$ (diamonds). Dashed reference line at nominal 95\%.}
  \label{fig:8}
\end{figure}

Figure~\ref{fig:8} presents 95\% coverage rates for the difference-in-differences in trend under AR[3] error structures. PW maintained near-nominal coverage across all conditions and series lengths, ranging from $\sim$90--93\% throughout. In contrast, OLS-NW exhibited substantial undercoverage across all three AR[3] conditions, with deficits generally larger than those observed under AR[2]. Under mild positive autocorrelation, coverage remained persistently low at $\sim$65--70\% across all series lengths and effect sizes, with no meaningful improvement as the number of periods increased. Under oscillatory
autocorrelation, coverage began at $\sim$67\% at short series lengths and improved gradually to $\sim$83--85\% by 100 periods, although it did not approach nominal levels within the range examined. Under high persistent positive autocorrelation, OLS-NW exhibited the most severe pattern: coverage began near 80\% at short series lengths and declined monotonically and steeply, falling to $\sim$45\% at 100 periods. This progressive deterioration with increasing sample size is more pronounced than under AR[2] and reinforces the conclusion that the inferential performance of OLS-NW worsens as both series length and AR order increase under persistent autocorrelation.

\subsubsection{Type I Error}

\begin{figure}[H]
  \centering
  \includegraphics[width=\textwidth, height=0.35\textheight, keepaspectratio]{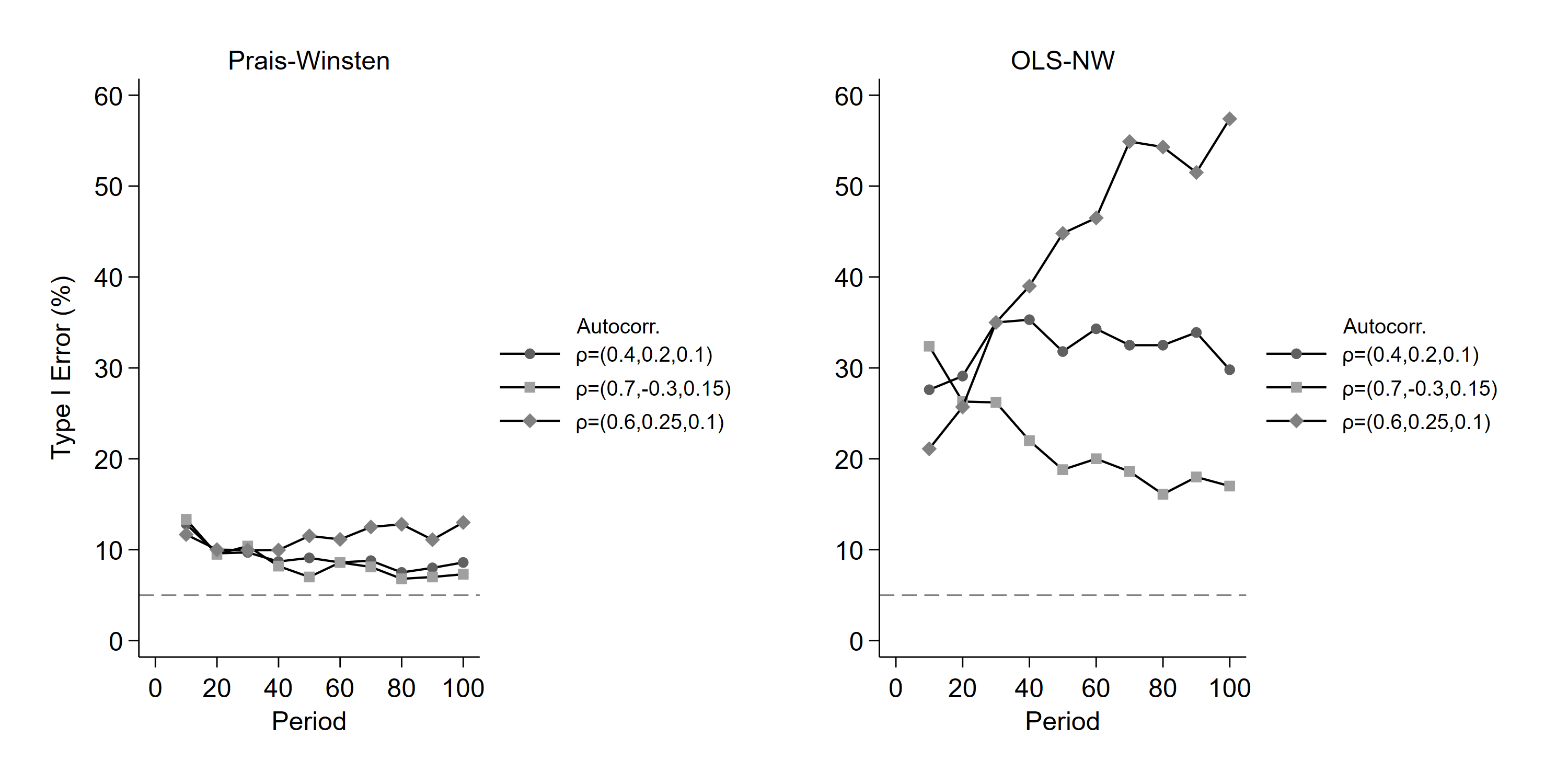}
  \caption{Type~I error rates for the difference-in-differences in trend under AR[3] error structures. Left panel: Prais-Winsten; right panel: OLS-NW. Lines distinguish autocorrelation scenarios: mild positive $\rho=(0.4, 0.2, 0.1)$ (circles); oscillatory $\rho=(0.7, -0.3, 0.15)$ (squares); high persistent $\rho=(0.6, 0.25, 0.1)$ (diamonds). Dashed reference line at nominal 5\%.}
  \label{fig:9}
\end{figure}

Figure~\ref{fig:9} presents Type~I error rates for the difference-in-differences in trend under AR[3] error structures. The patterns mirror those under AR[2] but with greater severity. PW remained modestly elevated at $\sim$10--13\%. OLS-NW was markedly inflated across all conditions: $\sim$28--35\% under mild positive autocorrelation (stable), declining from $\sim$32\% to $\sim$18\% under oscillatory autocorrelation, and rising monotonically from $\sim$21\% to $\sim$57\% under high persistent autocorrelation.

\subsubsection{Bias}

\begin{figure}[H]
  \centering
  \includegraphics[width=0.85\textwidth, height=0.85\textheight, keepaspectratio]{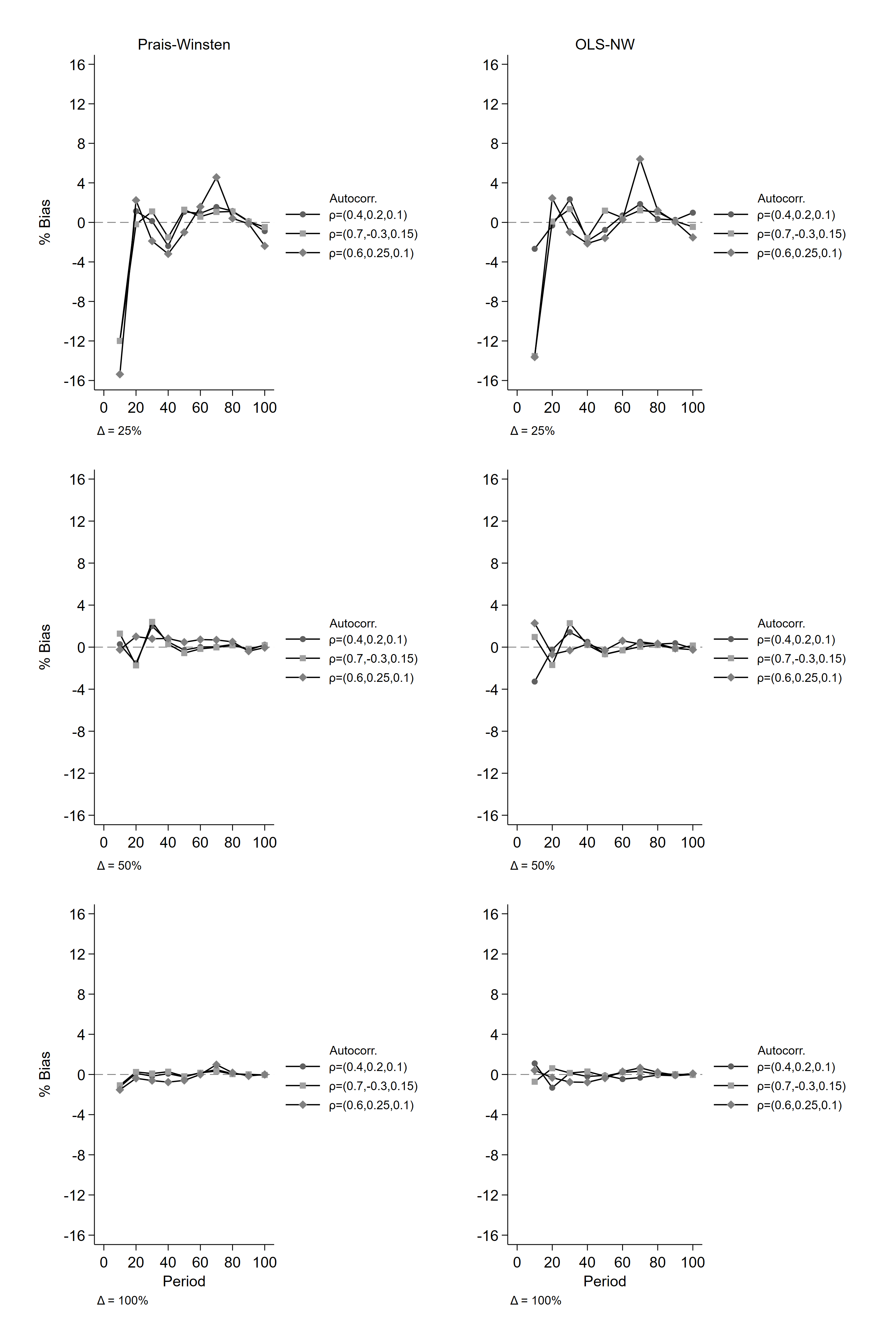}
  \caption{Percentage bias for the difference-in-differences in trend under AR[3] error structures. Left column: Prais-Winsten; right column: OLS-NW. Rows represent effect sizes (25\%, 50\%, 100\%). Lines distinguish autocorrelation scenarios: mild positive $\rho=(0.4, 0.2, 0.1)$ (circles); oscillatory $\rho=(0.7, -0.3, 0.15)$ (squares); high persistent $\rho=(0.6, 0.25, 0.1)$ (diamonds). Dashed reference line at zero.}
  \label{fig:10}
\end{figure}

Figure~\ref{fig:10} presents percentage bias under AR[3]. Bias patterns mirrored those observed under AR[2]: both methods remained essentially unbiased throughout, with only minor instability at very short series lengths under high persistent autocorrelation.

\subsubsection{RMSE}

\begin{figure}[H]
  \centering
  \includegraphics[width=0.85\textwidth, height=0.85\textheight, keepaspectratio]{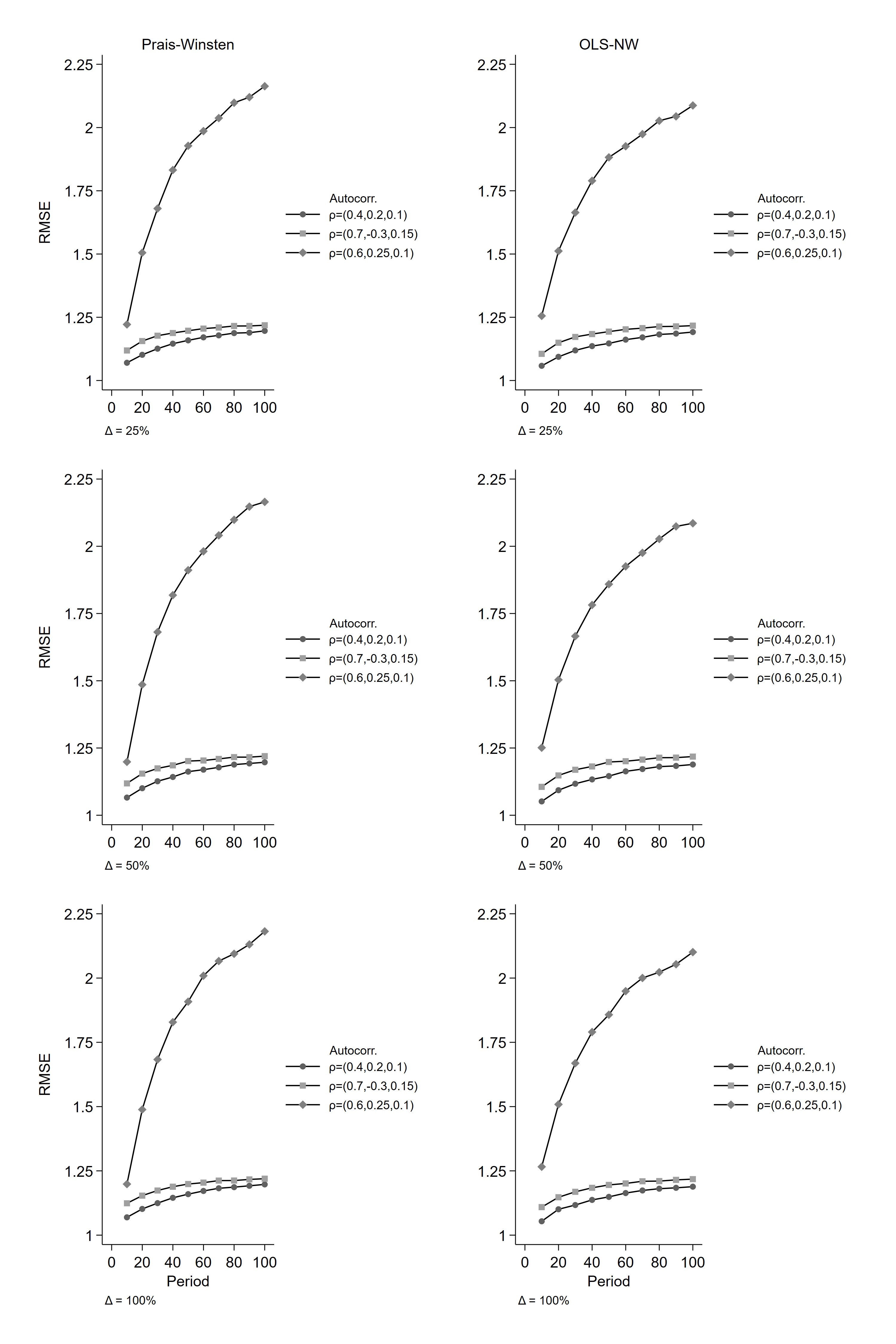}
  \caption{Root mean squared error (RMSE) for the difference-in-differences in trend under AR[3] error structures. Left column: Prais-Winsten; right column: OLS-NW. Rows represent effect sizes (25\%, 50\%, 100\%). Lines distinguish autocorrelation scenarios: mild positive $\rho=(0.4, 0.2, 0.1)$ (circles); oscillatory $\rho=(0.7, -0.3, 0.15)$ (squares); high persistent $\rho=(0.6, 0.25, 0.1)$ (diamonds).}
  \label{fig:11}
\end{figure}

Figure~\ref{fig:11} presents RMSE under AR[3]. The pattern was consistent with AR[2]: low and stable under mild positive and oscillatory autocorrelation ($\sim$1.0--1.2), rising monotonically under high persistent autocorrelation to $\sim$2.1--2.2 at 100 periods --- somewhat higher than the AR[2] values, reflecting greater variance at higher AR order. Differences between methods were modest throughout.

\subsubsection{Empirical Standard Errors}

\begin{figure}[H]
  \centering
  \includegraphics[width=0.85\textwidth, height=0.85\textheight, keepaspectratio]{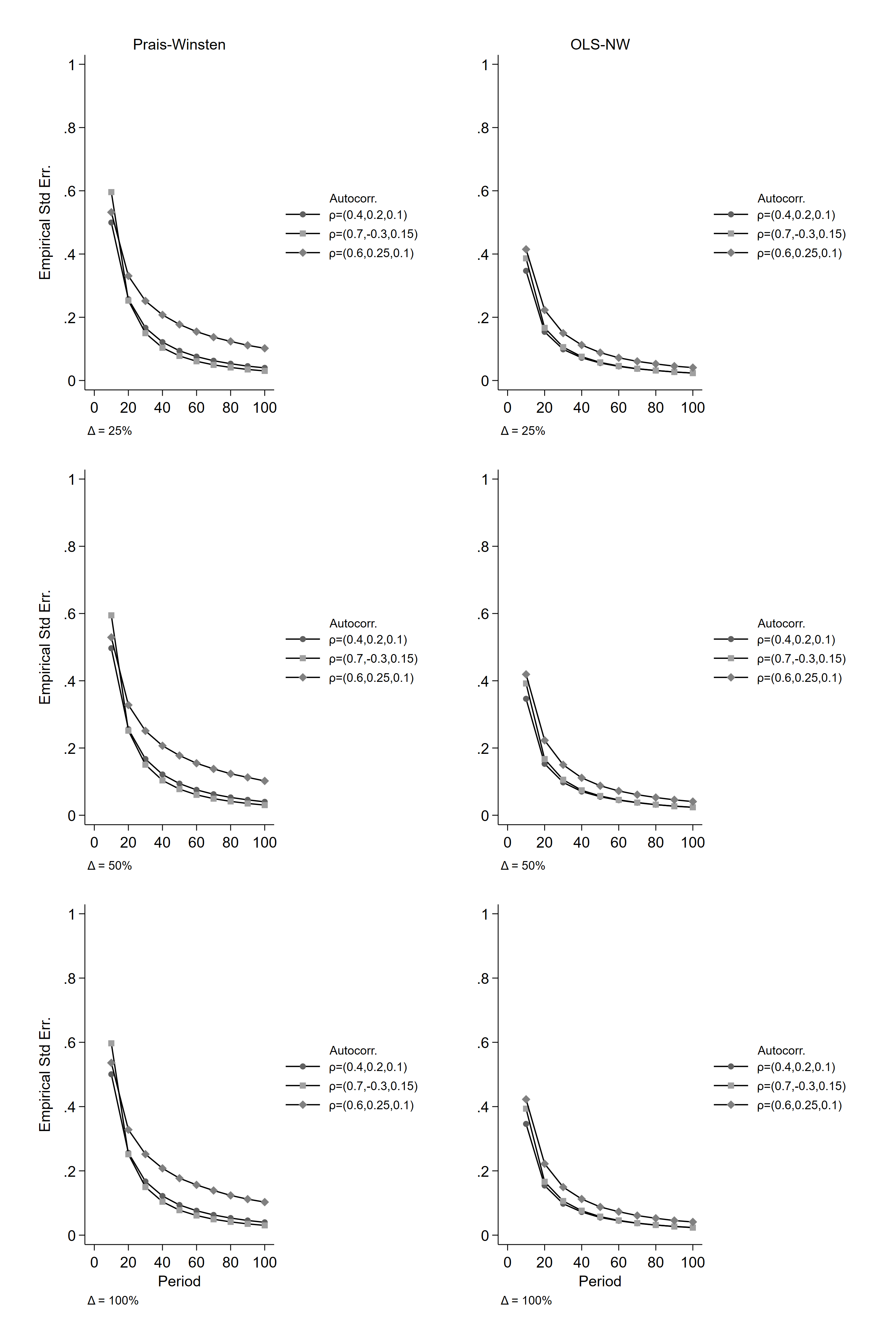}
  \caption{Empirical standard errors for the difference-in-differences in trend under AR[3] error structures. Left column: Prais-Winsten; right column: OLS-NW. Rows represent effect sizes (25\%, 50\%, 100\%). Lines distinguish autocorrelation scenarios: mild positive $\rho=(0.4, 0.2, 0.1)$ (circles); oscillatory $\rho=(0.7, -0.3, 0.15)$ (squares); high persistent $\rho=(0.6, 0.25, 0.1)$ (diamonds).}
  \label{fig:12}
\end{figure}

Figure~\ref{fig:12} presents empirical standard errors under AR[3]. The pattern was consistent with AR[2]: PW produced larger standard errors throughout, with the gap narrowing under mild positive and oscillatory autocorrelation and persisting under high persistent autocorrelation. The same trade-off applies: OLS-NW's smaller standard errors reflect inferential inflation, not efficiency.

\subsection{Difference-in-Differences in Level: AR{[2]}}

Level change results are presented in Appendix~I (Figures~1--6 for AR[2] and
Figures~7--12 for AR[3]). Key findings are summarized here.

Under AR[2], neither method reached maximum power within the 100-period range under most conditions. The relative performance of OLS-NW and PW varied by autocorrelation scenario: OLS-NW outperformed PW under mild positive autocorrelation, consistent with the same anticonservative behavior seen in the trend results. PW outperformed OLS-NW under oscillatory autocorrelation at shorter series lengths, and under high persistent autocorrelation, both methods plateaued well below maximum power across all effect sizes, with the plateau ranging from $\sim$45--50\% at the smallest effect size to $\sim$65--70\% at the largest, and minimal separation between methods throughout. Coverage and Type~I error patterns broadly mirrored the trend results, with one notable difference: under oscillatory autocorrelation, OLS-NW Type~I error fell below the nominal 5\% level at longer series lengths, reflecting an overly conservative inferential behavior that directly corresponds to PW's power advantage in that condition. Bias remained negligible throughout, and RMSE patterns closely mirrored those for trend change. Empirical standard errors revealed a notable reversal under oscillatory autocorrelation, where PW eventually produced smaller standard errors than OLS-NW at longer series lengths, consistent with PW's power advantage under that condition.

\subsection{Difference-in-Differences in Level: AR{[3]}}

Under AR[3], neither method reached maximum power under most conditions, and the relative performance of the two methods was more variable than under AR[2]. OLS-NW outperformed PW under mild positive autocorrelation, reflecting the same anticonservative behavior documented in the trend results. Under oscillatory autocorrelation, neither method showed a consistent advantage, with power profiles crossing at various series lengths. Under high persistent autocorrelation, both methods remained substantially underpowered across all effect sizes. PW demonstrated a power advantage over OLS-NW in two settings: under oscillatory autocorrelation for level change at both AR[2] and AR[3], and under high persistent autocorrelation at the largest effect size (30\%) for AR[3] level change. Coverage and Type~I error patterns were broadly consistent with the trend AR[3] results, with somewhat less extreme deterioration under high persistent autocorrelation. Bias remained negligible and RMSE patterns were similar to the trend results. Under high persistent autocorrelation, a reversal in empirical standard errors was observed, with PW eventually producing smaller standard errors than OLS-NW, consistent with PW's power advantage in that condition.

\subsection{Sensitivity Analyses}

Appendix~II presents sensitivity analysis results for AR[2] (Figure~1) and AR[3] (Figure~2). Both were fully consistent with the primary findings: OLS-NW achieved higher power at shorter series lengths, PW maintained near-nominal coverage, OLS-NW coverage deteriorated progressively under high persistent autocorrelation, bias was negligible for both methods, and RMSE was elevated under high persistent autocorrelation. The primary results were not sensitive to the alternative design specifications.

\subsection{Misspecification Analysis}

Appendix~III, Figures~1 and 2 present coverage and Type~I error results for the misspecification analysis.

For PW, the correct AR[2] and underspecified AR[1] lines were nearly
indistinguishable across all three autocorrelation scenarios and the full range of series lengths. Under mild positive autocorrelation, both specification conditions produced modestly elevated but stable Type~I error rates of $\sim$7--12\%, tracking each other closely throughout. Under oscillatory autocorrelation, both conditions produced near-nominal Type~I error, with PW becoming slightly conservative at longer series lengths under both specification conditions. Under high persistent autocorrelation, both conditions produced modestly elevated Type~I error of
$\sim$10--13\%, with no meaningful separation between them.

For OLS-NW, the correct AR[2] and underspecified AR[1] lines were similarly
indistinguishable. Under mild positive autocorrelation, both conditions produced stable Type~I error of $\sim$25--30\%, with the two lines tracking in
parallel throughout. Under oscillatory autocorrelation, both conditions started elevated and declined toward nominal at longer series lengths, with near-identical trajectories. Under high persistent autocorrelation, both conditions exhibited a monotonic increase in Type~I error with series length, reaching $\sim$50--60\% at 100 periods, with the correct and underspecified lines remaining essentially equivalent throughout.

\section{Applied Example}

\subsection{Background and Study Design}

To illustrate the practical consequences of estimator choice under higher-order autoregressive errors, we analyze an artificial dataset reflecting a realistic disease management study. Disease management programs target individuals with chronic conditions or elevated risk through structured behavioral and clinical interventions \citep{linden2008,kullgren2018}. Prediabetes is a common target given the effectiveness of lifestyle interventions in preventing progression to type 2 diabetes \citep{biuso2007}.

The artificial study involves five medical group practices whose patients with pre-diabetes were fitted with continuous glucose monitors (CGMs) and tracked for 180 days to establish a daily baseline average blood glucose level. One practice was then randomly assigned to implement a comprehensive lifestyle intervention; the remaining four served as controls. All practices were monitored for a further 180 days following the intervention. The outcome was the daily average blood glucose level (mg/dL) aggregated across patients within each practice.

Data were generated using the \pkg{itsadgp} package \citep{linden2024}. Baseline intercepts were set to 108 mg/dL for all practices, within the clinical pre-diabetes range of 100--125 mg/dL. The pre-intervention trend was +0.05 mg/dL per day for all groups, reflecting a gradual increase in blood glucose levels consistent with untreated prediabetes. Control practices maintained this trend throughout the study, with average blood glucose approaching $\sim$117 mg/dL by day 360. The treated practice's post-intervention trend was set to $-$0.03 mg/dL per day, reflecting modest improvement following the intervention. No immediate level change was specified at the intervention point. The true difference-in-differences in trend was therefore $(-0.03 - 0.05) - (0.05 - 0.05) = -0.08$ mg/dL per day. A standard deviation of 3 mg/dL was applied to reflect realistic day-to-day glucose variability. High persistent positive autocorrelation was specified to reflect plausible serial dependence in longitudinal glucose measurements. Three datasets were generated using identical parameters and a common random seed (77,777), differing only in AR order: AR[1] ($\rho = 0.7$), AR[2]
($\rho = (0.7, 0.2)$), and AR[3] ($\rho = (0.6, 0.25, 0.1)$). Each dataset
was analyzed using both OLS-NW and Prais--Winsten regression, with the lag
order matched to the AR order of the generating process.

\subsection{Results}

Table~\ref{tab:applied} presents the difference-in-differences in trend
estimates, standard errors, 95\% confidence intervals, and $p$-values for each
AR order and estimation method. These results represent a single realization;
coefficient estimates will vary across replications, though the simulation study
confirms both methods are approximately unbiased on average. The pattern that
remains stable is the growing standard error inflation under OLS-NW relative to
PW as AR order increases --- consistent with the simulation findings.

Under AR[1], both methods agreed: the intervention significantly reduced the rate
of glucose increase, with similar coefficients and the same clinical conclusion.
Under AR[2], both remained significant but OLS-NW's standard error was nearly
three times smaller than PW's, reflecting the anticonservative inference
documented in the simulations. Under AR[3], the methods reached opposite
conclusions --- OLS-NW highly significant ($p < 0.001$), PW non-significant
($p = 0.242$) with a confidence interval that included zero. A disease management
intervention could be recommended for broad implementation based on OLS-NW's
false confidence, while PW correctly signals insufficient evidence.

\section{Discussion}

The fundamental finding of this study is that the trade-off between power and inferential validity documented under AR[1] \citep{linden2026a} not only persists under higher-order autoregressive errors but intensifies in ways that depend critically on the autocorrelation structure. The two methods are not simply better or worse versions of each other. They handle serial dependence through fundamentally different mechanisms, and the consequences of those differences change as the error process becomes more complex. Table~\ref{tab:summary} summarizes performance across AR orders; AR[1] results are from Linden \citep{linden2026a}.

\subsection{Statistical Power}

OLS-NW's higher power is not a sign of statistical efficiency. It is a consequence of the HAC estimator's tendency to underestimate the true long-run variance under positive autocorrelation, producing standard errors that are too small and rejection rates that are too high \citep{wooldridge2020}. This distinction matters because a researcher who chooses OLS-NW for its power advantage is, in practice, accepting an elevated false positive rate as the price. Under high persistent autocorrelation, the underestimation grows with both AR order and series length, so the apparent power advantage widens precisely when it is least trustworthy.

The exception is informative. Under oscillatory autocorrelation, the NW estimator overcorrects for level change, producing inflated standard errors; PW avoids this and achieves a genuine power advantage. The reversal shows that the relationship between the two methods depends on how well the bandwidth selection procedure approximates the spectral structure of the errors.

\subsection{Coverage and Type~I Error}

The coverage and Type~I error results reveal a more troubling problem than simple inflation. OLS-NW's inferential behavior is qualitatively different depending on the autocorrelation structure, and in one important case it deteriorates as series length increases rather than improving.

Under mild positive autocorrelation, the HAC estimator's underestimation of the long-run variance is roughly stable, producing persistent but bounded Type~I error inflation. A researcher using OLS-NW in this setting will consistently reject too often, but by a predictable margin. Under high persistent autocorrelation, the situation is fundamentally different: the long-run variance grows faster than the bandwidth can track, so the underestimation widens with series length. Coverage falls and Type~I error rises with series length, the opposite of what asymptotic theory predicts, and a pattern with no counterpart under AR[1] \citep{linden2026a}. For practitioners, this means that collecting more data does not improve OLS-NW's inferential reliability under persistent higher-order autocorrelation.

PW's modest Type~I error elevation at AR[3] ($\sim$10--13\%) reflects the finite-sample cost of estimating additional Yule--Walker parameters, not a structural deficiency. Unlike OLS-NW's inflation, PW's elevation does not grow with series length and remains well within the range of acceptable inferential practice.

\subsection{Bias, Standard Errors, and RMSE}

Both methods are essentially unbiased. The key implication is that differences in coverage and power are not attributable to biased point estimates but to how each method handles variance. PW's larger standard errors reflect the known efficiency cost of FGLS estimation \citep{wooldridge2020}; they are honest estimates of uncertainty. OLS-NW's smaller standard errors under most conditions reflect underestimation of the true long-run variance; they create false precision.

The standard error crossings under oscillatory and high persistent autocorrelation for level change are a direct expression of the same mechanism. When the NW bandwidth overestimates the long-run variance, OLS-NW's standard errors become inflated, and PW's efficient estimates become relatively smaller. The crossing is not a peculiarity of the simulation design; it is a predictable consequence of how each estimator responds to these specific autocorrelation structures.

Under high persistent autocorrelation, RMSE rises monotonically with series length for both methods, a reminder that statistical difficulties in this setting are not resolved by larger samples. The variance of the AR process itself grows as the series lengthens, compounding the inferential problems already documented for OLS-NW.

\subsection{Practical Recommendations}

Characterizing the autocorrelation structure before selecting an estimator is essential. Under mild positive autocorrelation, the choice involves a familiar trade-off: PW for valid inference, OLS-NW only if power is paramount and inflated Type~I error is explicitly acknowledged. Under oscillatory autocorrelation, PW is the clear choice in all settings; OLS-NW's behavior is erratic and direction-dependent, making it difficult to interpret or justify. Under high persistent autocorrelation, OLS-NW cannot be recommended at any series length. Its inferential deficiencies worsen as the series grows, removing the usual asymptotic justification entirely. PW is the only defensible option.

The misspecification results add a practical note of reassurance for PW users: fitting an AR[1] model when the true process is AR[2] does not meaningfully degrade PW's inferential performance. The GLS transformation is robust to moderate lag order misspecification. For OLS-NW, the converse is also true but less encouraging: fitting a higher-order AR model will not fix the inferential problems. The HAC estimator's deficiencies are structural, not a consequence of lag order choice. Regardless of estimator, researchers should assess autocorrelation structure, report design features, and conduct sensitivity analyses \citep{linden2005}.

\subsection{Limitations}

The simulation design reflects a common MG-ITSA application: a single treated unit, four controls, a single intervention point, and regression-based estimation. Several factors not examined here may affect performance, including multiple treatments, seasonality, and non-Gaussian outcomes. When seasonality is present but cannot be modeled directly, longer series are generally required \citep{hyndman2007}. The impact of varying the number of control units under higher-order autoregressive structures remains an open question. The maximum series length was 100 periods, and some conditions under high persistent autocorrelation had not stabilized. Even longer series may be required to fully characterize asymptotic behavior.

Several limitations are specific to the higher-order autoregressive setting. The autocorrelation scenarios examined represent qualitatively distinct patterns but are not exhaustive. Other parameterizations, particularly those nearer the boundary of stationarity, may yield different performance profiles. The modest elevation in PW Type~I error at AR[3] likely reflects finite-sample limitations of Yule--Walker estimation with multiple parameters; whether this elevation grows further at AR[4] and beyond is unknown. Finally, only OLS-NW and PW were evaluated. While a broader range of methods has been proposed for ITS analysis \citep{ewusie2020}, most are designed for single-group settings and are not directly applicable to the MG-ITSA between-group contrast structure examined here. Future work should consider alternative approaches compatible with the MG-ITSA design, including ARCH-family models \citep{harvey1989,enders2004}, Bayesian frameworks \citep{ma2025}, instrumental variables \citep{linden2006}, and machine learning algorithms \citep{linden2018b}.

\section{Conclusion}

The trade-off between power and inferential validity documented under AR[1] \citep{linden2026a} persists and intensifies under higher-order autoregressive errors in the MG-ITSA design. OLS-NW's apparent power advantage reflects anticonservative inference rather than genuine statistical efficiency, and its inferential deficiencies worsen with AR order and, under high persistent autocorrelation, with series length. PW provided well-calibrated inference across virtually all conditions and is the preferred estimator when valid hypothesis testing is the priority. Both methods were essentially unbiased, confirming that the differences between them are about variance estimation, not point accuracy. Future research should examine estimator performance under higher-order serial dependence structures for non-Gaussian outcomes and multiple intervention designs.


\newpage
\bibliography{references}

\begin{thebibliography}{10}

\bibitem{campbell1966}
Campbell DT, Stanley JC.
\newblock Experimental and Quasi-Experimental Designs for Research.
\newblock Chicago: Rand McNally; 1966.

\bibitem{shadish2002}
Shadish WR, Cook TD, Campbell DT.
\newblock Experimental and Quasi-Experimental Designs for Generalized Causal Inference.
\newblock Boston: Houghton Mifflin; 2002.

\bibitem{linden2013}
Linden A.
\newblock Assessing regression to the mean effects in health care initiatives.
\newblock BMC Medical Research Methodology. 2013;13:119.
\newblock Available from: \url{https://doi.org/10.1186/1471-2288-13-119}.

\bibitem{linden2016}
Linden A, Yarnold PR.
\newblock Using machine learning to identify structural breaks in single-group interrupted time series designs.
\newblock Journal of Evaluation in Clinical Practice. 2016;22:855-9.
\newblock Available from: \url{https://doi.org/10.1111/jep.12544}.

\bibitem{linden2015}
Linden A.
\newblock Conducting interrupted time-series analysis for single- and multiple-group comparisons.
\newblock Stata Journal. 2015;15(2):480-500.
\newblock Available from: \url{https://doi.org/10.1177/1536867X1501500208}.

\bibitem{linden2017a}
Linden A.
\newblock Challenges to validity in single-group interrupted time series analysis.
\newblock Journal of Evaluation in Clinical Practice. 2017;23:413-8.
\newblock Available from: \url{https://doi.org/10.1111/jep.12638}.

\bibitem{linden2017b}
Linden A.
\newblock Persistent threats to validity in single-group interrupted time series analysis with a crossover design.
\newblock Journal of Evaluation in Clinical Practice. 2017;23:419-25.
\newblock Available from: \url{https://doi.org/10.1111/jep.12668}.

\bibitem{abadie2010}
Abadie A, Diamond A, Hainmueller J.
\newblock Synthetic control methods for comparative case studies: Estimating the effect of {California}'s tobacco control program.
\newblock Journal of the American Statistical Association. 2010;105(490):493-505.
\newblock Available from: \url{https://doi.org/10.1198/jasa.2009.ap08746}.

\bibitem{linden2018}
Linden A.
\newblock A matching framework to improve causal inference in interrupted time-series analysis.
\newblock Journal of Evaluation in Clinical Practice. 2018;24:408-15.
\newblock Available from: \url{https://doi.org/10.1111/jep.12874}.

\bibitem{rubin1974}
Rubin DB.
\newblock Estimating causal effects of treatments in randomized and non-randomized studies.
\newblock Journal of Educational Psychology. 1974;66:688-701.

\bibitem{linden2017c}
Linden A.
\newblock A comprehensive set of postestimation measures to enrich interrupted time-series analysis.
\newblock Stata Journal. 2017;17:73-88.
\newblock Available from: \url{https://doi.org/10.1177/1536867X1501500208}.

\bibitem{newey1987}
Newey WK, West KD.
\newblock A simple, positive semi-definite, heteroskedasticity and autocorrelation consistent covariance matrix.
\newblock Econometrica. 1987;55:703-8.
\newblock Available from: \url{https://doi.org/10.2307/1913610}.

\bibitem{prais1954}
Prais SJ, Winsten CB.
\newblock Trend estimators and serial correlation.
\newblock Cowles Commission; 1954. 383.
\newblock Available from: \url{https://cowles.yale.edu/sites/default/files/files/pub/cdp/s-0383.pdf}.

\bibitem{harvey1989}
Harvey AC.
\newblock Forecasting, Structural Time Series Models and the {Kalman} Filter.
\newblock Cambridge: Cambridge University Press; 1989.

\bibitem{enders2004}
Enders W.
\newblock Applied Econometric Time Series.
\newblock 2nd ed. Hoboken, NJ: Wiley; 2004.

\bibitem{turner2021}
Turner SL, Forbes AB, Karahalios A, Taljaard M, McKenzie JE.
\newblock Evaluation of statistical methods used in the analysis of interrupted time series studies: a simulation study.
\newblock BMC Medical Research Methodology. 2021;21:181.
\newblock Available from: \url{https://doi.org/10.1186/s12874-021-01393-1}.

\bibitem{bottomley2023}
Bottomley C, Ooko M, Gasparrini A, Keogh RH.
\newblock In praise of {Prais--Winsten}: an evaluation of methods used to account for autocorrelation in interrupted time series.
\newblock Statistics in Medicine. 2023;42(8):1277-88.
\newblock Available from: \url{https://doi.org/10.1002/sim.9547}.

\bibitem{linden2026a}
Linden A. Adjustment for autocorrelation in multiple-group (controlled) interrupted time series analysis and its effect on power: a simulation study of the {Newey-West} and {Prais-Winsten} methods; 2026.
\newblock Preprint. Research Square.
\newblock Available from: \url{https://doi.org/10.21203/rs.3.rs-8865851/v1}.

\bibitem{vougas2021}
Vougas DV.
\newblock {Prais--Winsten} algorithm for regression with second or higher order autoregressive errors.
\newblock Econometrics. 2021;9(3):32.
\newblock Available from: \url{https://doi.org/10.3390/econometrics9030032}.

\bibitem{linden2026b}
Linden A. {PRAISK}: {Stata} module for computing iterated {Prais-Winsten} regression with {AR($k$)} errors; 2026.
\newblock Statistical Software Components s459648, Boston College Department of Economics.

\bibitem{kutner2005}
Kutner MH, Nachtsheim CJ, Neter J, Li W.
\newblock Applied Linear Statistical Models.
\newblock 5th ed. New York: McGraw-Hill Irwin; 2005.

\bibitem{linden2022}
Linden A.
\newblock Erratum: a comprehensive set of postestimation measures to enrich interrupted time-series analysis.
\newblock Stata Journal. 2022;22:231-3.
\newblock Available from: \url{https://doi.org/10.1177/1536867X221083929}.

\bibitem{wooldridge2020}
Wooldridge JM.
\newblock Introductory Econometrics: A Modern Approach.
\newblock 7th ed. Boston: Cengage; 2020.

\bibitem{newey1994}
Newey WK, West KD.
\newblock Automatic lag selection in covariance matrix estimation.
\newblock Review of Economic Studies. 1994;61(4):631-53.

\bibitem{judge1985}
Judge GG, Griffiths WE, Hill RC, L{\"u}tkepohl H, Lee TC.
\newblock The Theory and Practice of Econometrics.
\newblock 2nd ed. Wiley; 1985.

\bibitem{hamilton1994}
Hamilton JD.
\newblock Time Series Analysis.
\newblock Princeton University Press; 1994.

\bibitem{golub1996}
Golub GH, Van~Loan CF.
\newblock Matrix Computations.
\newblock 3rd ed. Johns Hopkins University Press; 1996.

\bibitem{galbraith1974}
Galbraith RF, Galbraith JI.
\newblock On the inverses of some patterned matrices arising in the theory of stationary time series.
\newblock Journal of Applied Probability. 1974;11(1):63-71.

\bibitem{park1980}
Park RE, Mitchell BM.
\newblock Estimating the autocorrelated error model with trended data.
\newblock Journal of Econometrics. 1980;13(2):185-201.

\bibitem{brockwell1991}
Brockwell PJ, Davis RA.
\newblock Time Series: Theory and Methods.
\newblock 2nd ed. Springer; 1991.

\bibitem{linden2025}
Linden A. {POWER\_ITSA}: {Stata} module to compute power for single and multiple-group interrupted time series analysis; 2025.
\newblock Statistical Software Components S459461, Boston College Department of Economics.

\bibitem{linden2024}
Linden A. {ITSADGP}: {Stata} module to generate artificial data for interrupted time-series analysis; 2024.
\newblock Statistical Software Components S459403, Boston College Department of Economics.

\bibitem{linden2026c}
Linden A.
\newblock Power considerations for multiple-group (controlled) interrupted time series analysis: a comprehensive simulation study.
\newblock Evaluation \& the Health Professions. 2026.
\newblock Epub ahead of print.
\newblock Available from: \url{https://doi.org/10.1177/01632787261428159}.

\bibitem{burton2006}
Burton A, Altman DG, Royston P, Holder RL.
\newblock The design of simulation studies in medical statistics.
\newblock Statistics in Medicine. 2006;25:4279-92.
\newblock Available from: \url{https://doi.org/10.1002/sim.2673}.

\bibitem{linden2008}
Linden A, Adler-Milstein J.
\newblock Medicare disease management in policy context.
\newblock Health Care Financing Review. 2008;29:1-11.

\bibitem{kullgren2018}
Kullgren JT, Krupka E, Schachter A, Linden A, Miller J, Acharya Y, et~al.
\newblock Precommitting to choose wisely about low-value services: a stepped wedge cluster randomised trial.
\newblock BMJ Quality and Safety. 2018;27:355-64.
\newblock Available from: \url{https://doi.org/10.1136/bmjqs-2017-006699}.

\bibitem{biuso2007}
Biuso TJ, Butterworth S, Linden A.
\newblock A conceptual framework for targeting prediabetes with lifestyle, clinical and behavioral management interventions.
\newblock Disease Management. 2007;10(1):6-15.
\newblock Available from: \url{https://doi.org/10.1089/dis.2006.628}.

\bibitem{linden2005}
Linden A, Roberts N.
\newblock A user's guide to the disease management literature: recommendations for reporting and assessing program outcomes.
\newblock American Journal of Managed Care. 2005;11:113-20.

\bibitem{hyndman2007}
Hyndman RJ, Kostenko AV.
\newblock Minimum sample size requirements for seasonal forecasting models.
\newblock Foresight. 2007;6:12-5.

\bibitem{ewusie2020}
Ewusie JE, Soobiah C, Blondal E, Beyene J, Thabane L, Hamid JS.
\newblock Methods, applications and challenges in the analysis of interrupted time series data: a scoping review.
\newblock Journal of Multidisciplinary Healthcare. 2020;13:411-23.
\newblock Available from: \url{https://doi.org/10.2147/JMDH.S241085}.

\bibitem{ma2025}
Ma Y, Benmarhnia T.
\newblock Interrupted time series analysis in environmental epidemiology: a review of traditional and novel modeling approaches.
\newblock Current Environmental Health Reports. 2025;12:50.
\newblock Available from: \url{https://doi.org/10.1007/s40572-025-00517-3}.

\bibitem{linden2006}
Linden A, Adams JL.
\newblock Evaluating disease management programme effectiveness: an introduction to instrumental variables.
\newblock Journal of Evaluation in Clinical Practice. 2006;12:148-54.
\newblock Available from: \url{https://doi.org/10.1111/j.1365-2753.2006.00615.x}.

\bibitem{linden2018b}
Linden A, Yarnold PR.
\newblock Using machine learning to evaluate treatment effects in multiple-group interrupted time series analysis.
\newblock Journal of Evaluation in Clinical Practice. 2018;24:740-4.
\newblock Available from: \url{https://doi.org/10.1111/jep.12966}.

\end{thebibliography}


\newpage

\section*{Abbreviations}

AR: Autoregressive; FGLS: Feasible generalized least squares; GLS: Generalized
least squares; HAC: Heteroskedasticity- and autocorrelation-consistent standard
errors; ITSA: Interrupted time series analysis; MG-ITSA: Multiple-group
interrupted time series analysis; OLS: Ordinary least squares; OLS-NW: Ordinary
least squares with Newey-West standard errors; PW: Prais-Winsten regression;
RMSE: Root mean squared error; SG-ITSA: Single-group interrupted time series
analysis.

\section*{Supplementary Information}

The Supplement contains figures for difference-in-differences in level (Appendix~I), sensitivity analyses (Appendix~II), and misspecification analysis (Appendix~III). Stata code used in this paper is found at: \url{https://github.com/ariellinden/MG_ITSA_higher_order_ar_errors}

\section*{Authors' Contributions}

AL conceived the study and its design, conducted all analyses, wrote the
manuscript, and takes public responsibility for its content.

\section*{Funding}

There was no funding associated with this work.

\section*{Ethics Approval and Consent to Participate}

Not applicable.

\section*{Consent for Publication}

Not applicable.

\section*{Competing Interests}

The author declares no competing interests.

\section*{Acknowledgements}

I am grateful to Dimitrios V. Vougas for graciously providing the MATLAB code
that served as the basis for implementation of the \pkg{praisk} package.


\clearpage

\begin{singlespace}
\begin{table}[H]
\caption{Simulation inputs.}
\label{tab:sim_inputs}
\centering
\small
\begin{tabular}{>{\raggedright\arraybackslash}p{6.5cm}
                >{\centering\arraybackslash}p{2.5cm}
                >{\raggedright\arraybackslash}p{7.0cm}}
\toprule
\textbf{Parameter} & \textbf{Equation (1) term} & \textbf{Input values} \\
\midrule
Controls' baseline level
  & $\beta_0$
  & 10 \\[3pt]
Treated unit's baseline level
  & $\beta_0 + \beta_4$
  & 10 \\[3pt]
Controls' baseline trend
  & $\beta_1$
  & 1 \\[3pt]
Treated unit's baseline trend
  & $\beta_1 + \beta_5$
  & 1 \\[3pt]
Controls' level change 
  & $\beta_2$
  & 0 \\[3pt]
Treated unit's level change 
  & $\beta_2 + \beta_6$
  & 2, 2.5, 3 (representing 20\%, 25\%, and 30\%) \\[3pt]
Controls' post-treatment trend
  & $\beta_1 + \beta_3$
  & 1 \\[3pt]
Treated post-treatment trend
  & $\beta_1 + \beta_3 + \beta_5 + \beta_7$
  & 1.25, 1.50, 2 (representing 25\%, 50\%, and 100\%) \\[3pt]
Number of time periods
  &
  & 10--100 \\[3pt]
Initiation of treatment
  &
  & Halfway point \\[3pt]
Number of controls
  &
  & 4 \\[3pt]
Std. dev. for random error term
  &
  & 1 \\[6pt]
\multicolumn{3}{l}{\text{Autocorrelation coefficients:}} \\[4pt]
\quad AR{[2]}
  &
  & Scenario 1: $\rho = (0.4,\;0.2)$ \\
  & & Scenario 2: $\rho = (0.5,\;{-0.4})$ \\
  & & Scenario 3: $\rho = (0.7,\;0.2)$ \\[4pt]
\quad AR{[3]}
  &
  & Scenario 1: $\rho = (0.4,\;0.2,\;0.1)$ \\
  & & Scenario 2: $\rho = (0.7,\;{-0.3},\;0.15)$ \\
  & & Scenario 3: $\rho = (0.6,\;0.25,\;0.1)$ \\[6pt]
\multicolumn{3}{l}{\text{Misspecification analysis (AR[2] DGP only):}} \\[4pt]
\quad DGP AR order
  &
  & AR[2] \\[3pt]
\quad Fitted AR order
  &
  & AR[2] (correct); AR[1] (underspecified) \\[3pt]
\quad Autocorrelation scenarios
  &
  & Scenario 1: $\rho = (0.4,\;0.2)$ \\
  & & Scenario 2: $\rho = (0.5,\;{-0.4})$ \\
  & & Scenario 3: $\rho = (0.7,\;0.2)$ \\[3pt]
\quad Replications
  &
  & 2,000 \\
\bottomrule
\end{tabular}

\par\bigskip\noindent\raggedright
\footnotesize\textit{Note:} For the sensitivity analysis (combined level and
trend changes), the controls' baseline level is set to 8, the treatment unit's
level change is set to 2 (20\% increase), and the treatment unit's
post-treatment trend is set to 2 (100\% increase). All other values are
unchanged. For the misspecification analysis, the post-treatment effect (\texttt{tpost}) is set to 1.5 for power simulations and to 1 for Type~I error simulations; all other primary simulation inputs apply.
\end{table}
\end{singlespace}

\clearpage
\begin{singlespace}
\begin{table}[H]
\caption{Applied example results: difference-in-differences in trend
  (mg/dL per day) by AR order and estimation method. True effect $= -0.08$
  mg/dL per day.}
\label{tab:applied}
\centering
\small
\begin{tabular}{>{\raggedright\arraybackslash}p{1.8cm}
                >{\raggedright\arraybackslash}p{3.0cm}
                >{\centering\arraybackslash}p{2.2cm}
                >{\centering\arraybackslash}p{1.6cm}
                >{\centering\arraybackslash}p{1.6cm}
                >{\centering\arraybackslash}p{1.8cm}
                >{\centering\arraybackslash}p{1.8cm}}
\toprule
\textbf{AR Order} & \textbf{Method} & \textbf{Coefficient} & \textbf{SE} &
\textbf{$p$-value} & \multicolumn{2}{c}{\textbf{95\% CI}} \\
\cmidrule(lr){6-7}
 &  &  &  &  & \textbf{Lower} & \textbf{Upper} \\
\midrule
AR[1] & OLS-NW        & $-$0.0951 & 0.0106 & $<$0.001 & $-$0.1159 & $-$0.0743 \\[3pt]
      & Prais-Winsten & $-$0.0943 & 0.0224 & $<$0.001 & $-$0.1383 & $-$0.0503 \\[6pt]
AR[2] & OLS-NW        & $-$0.1241 & 0.0168 & $<$0.001 & $-$0.1571 & $-$0.0911 \\[3pt]
      & Prais-Winsten & $-$0.1216 & 0.0494 & 0.0139   & $-$0.2184 & $-$0.0247 \\[6pt]
AR[3] & OLS-NW        & $-$0.1430 & 0.0219 & $<$0.001 & $-$0.1859 & $-$0.1000 \\[3pt]
      & Prais-Winsten & $-$0.0919 & 0.0786 & 0.2423   & $-$0.2460 &    0.0622 \\
\bottomrule
\end{tabular}
\par\bigskip
\noindent\raggedright\footnotesize\textit{Note:} Both methods applied to the same dataset at each AR order. SE = standard error; CI = confidence interval.
\end{table}
\end{singlespace}
\clearpage

\begin{singlespace}
\setlength{\tabcolsep}{4pt}
\renewcommand{\arraystretch}{1.3}
\footnotesize
\captionsetup{justification=justified}
\captionof{table}{Summary comparison of simulation findings across AR[1], AR[2], and
AR[3] error structures for OLS-NW and Prais--Winsten regression in controlled
interrupted time series analysis. AR[1] results are from Linden
\citep{linden2026a}.}
\captionsetup{justification=raggedright}
\label{tab:summary}
\begin{longtable}{>{\raggedright\arraybackslash}p{2.5cm}
                  >{\raggedright\arraybackslash}p{4.5cm}
                  >{\raggedright\arraybackslash}p{4.5cm}
                  >{\raggedright\arraybackslash}p{4.5cm}}
\toprule
\textbf{Performance Measure} & \textbf{AR[1]} & \textbf{AR[2]} & \textbf{AR[3]} \\
\midrule
\endfirsthead
\multicolumn{4}{l}{\small\textit{Table~\ref{tab:summary} continued}}\\[4pt]
\toprule
\textbf{Performance Measure} & \textbf{AR[1]} & \textbf{AR[2]} & \textbf{AR[3]} \\
\midrule
\endhead
\midrule
\multicolumn{4}{r}{\small\textit{Continued on next page}}\\
\endfoot
\bottomrule
\endlastfoot

\textbf{Power}
  & OLS-NW consistently higher than PW across all conditions; differences often exceeding 10--15\%; however, this apparent advantage reflects anticonservative inference under positive autocorrelation rather than genuine gains in statistical efficiency
  & OLS-NW higher under mild positive and high persistent autocorrelation; PW higher for level change under oscillatory autocorrelation (driven by OLS-NW overcorrection); power gap widens under high persistent autocorrelation relative to AR[1]; OLS-NW's apparent advantage under mild and high persistent autocorrelation reflects inflated Type~I error rather than genuine efficiency gains
  & OLS-NW higher under mild positive and oscillatory autocorrelation; PW achieves a genuine power advantage for level change under oscillatory and high persistent autocorrelation at larger effect sizes; gap further widens relative to AR[2]; as at AR[2], OLS-NW's apparent power advantage under most conditions reflects anticonservative inference rather than genuine statistical efficiency \\[4pt]

\textbf{95\% Coverage}
  & PW maintains near-nominal coverage ($\sim$94\%); OLS-NW shows stable undercoverage ($\sim$82--83\%) across all conditions, worsening with more controls
  & PW maintains near-nominal coverage ($\sim$91--94\%); OLS-NW shows stable undercoverage ($\sim$75\%) under mild autocorrelation, and progressive monotonic deterioration to $\sim$50\% under high persistent autocorrelation---a pattern not observed at AR[1]
  & PW maintains near-nominal coverage ($\sim$91--93\%); OLS-NW undercoverage worsens to $\sim$65--70\% under mild autocorrelation and deteriorates even more steeply to $\sim$45--50\% under high persistent autocorrelation---more severe than AR[2] \\[4pt]

\textbf{Type~I Error}
  & PW near-nominal ($\sim$5\%); OLS-NW inflated ($\sim$18\% average; max $\sim$30\%), declining with increasing series length
  & PW near-nominal ($\sim$5\%); OLS-NW shows stable inflation ($\sim$25--30\%) under mild autocorrelation, partial decline under oscillatory autocorrelation, and monotonically increasing inflation reaching $\sim$50\% under high persistent autocorrelation---qualitatively different from AR[1]
  & PW modestly elevated ($\sim$10--13\%); OLS-NW shows stable inflation ($\sim$28--35\%) under mild autocorrelation, partial decline under oscillatory autocorrelation, and monotonically increasing inflation reaching $\sim$57\% under high persistent autocorrelation---more extreme than AR[2] \\[4pt]

\textbf{Bias}
  & Both methods approximately unbiased across all conditions; bias decreased with larger effect sizes and longer series
  & Both methods approximately unbiased across all conditions; minor instability at very short series under oscillatory and high persistent autocorrelation
  & Both methods approximately unbiased across all conditions and series lengths; minor instability at very short series under high persistent autocorrelation, similar to AR[2] \\[4pt]

\textbf{RMSE}
  & Approximately constant ($\sim$1.01) for both methods across all conditions with no meaningful difference between them
  & Low and flat ($\sim$1.0--1.15) for both methods under mild positive and oscillatory autocorrelation; rises substantially and monotonically to $\sim$1.9 at 100 periods under high persistent autocorrelation, with PW marginally higher than OLS-NW
  & Low and flat ($\sim$1.0--1.2) for both methods under mild positive and oscillatory autocorrelation; rises substantially and monotonically to $\sim$2.1--2.2 at 100 periods under high persistent autocorrelation---somewhat higher than AR[2] \\[4pt]

\textbf{Empirical Standard Errors}
  & PW consistently $\sim$25\% larger than OLS-NW across all conditions; both decline with increasing series length
  & PW larger than OLS-NW in most conditions; methods converge and cross under oscillatory autocorrelation for level change; PW becomes smaller than OLS-NW under high persistent autocorrelation for level change at longer series---consistent with PW's power advantage in that condition
  & PW larger than OLS-NW under mild positive and oscillatory autocorrelation with the gap persisting longer than at AR[2]; PW again smaller than OLS-NW under high persistent autocorrelation for level change at longer series---consistent with PW's power advantage in that condition \\[4pt]

\textbf{Sensitivity Analyses}
  & Consistent with main findings across all conditions
  & Consistent with main findings across all conditions and AR parameter configurations
  & Consistent with main findings across all conditions and AR parameter configurations; also consistent with AR[2] sensitivity findings \\

\end{longtable}
\end{singlespace}
\renewcommand{\arraystretch}{1}   


\clearpage

\end{document}